\documentclass[12pt,a4paper]{article}
\usepackage{graphics,graphicx}
\usepackage{amssymb,epsfig,amsmath,euscript,array}
\usepackage{cite}

\makeatletter
\@addtoreset{equation}{section}
\makeatother
\renewcommand{\theequation}{\thesection.\arabic{equation}}

\newcommand{\startappendix}{
\setcounter{section}{0}
\renewcommand{\thesection}{\Alph{section}}
\renewcommand{\theequation}{\Alph{section}.\arabic{equation}}}
\newcommand{\Appendix}[1]{
\refstepcounter{section}
\begin{flushleft}
{\Large\bf Appendix \thesection: #1}
\end{flushleft}}

\newcounter{multieqs}

\newcommand{\be}{\begin{equation}}
\newcommand{\ee}{\end{equation}}

\newcommand{\bm}[1]{\mbox{\boldmath $#1$}}

\newcommand{\kslash}{k \!\!\! / }

\newcommand{\lslash}{l \!\! / }
\newcommand{\Pslash}{P \!\!\!\! / }

\newcommand{\islash}{i \!\!\! / }
\newcommand{\jslash}{j \!\!\! / }
\newcommand{\aslash}{a \!\!\! / }
\newcommand{\bslash}{{b \hspace{-6pt} \slash} }

\newcommand{\onslash}{1 \!\!\! / }
\newcommand{\twslash}{2 \!\!\!/ }
\newcommand{\thslash}{3 \!\!\!/ }
\newcommand{\foslash}{4 \!\!\! / }
\newcommand{\fislash}{5 \!\!\! / }

\newcommand{\mslash}{m \!\!\! / }

\def\bd{\begin{document}}
\def\ed{\end{document}}
\def\nn{\nonumber}
\def\bea{\begin{eqnarray}}
\def\eea{\end{eqnarray}}
\def\eps{\epsilon}

\def\ab{(ijab)}
\def\ba{(ijba)}
\def\ijab{{\tr}_{+}(\islash\, \jslash\, \aslash \, \bslash)}
\def\ijba{{\tr}_{+}(\islash\, \jslash\, \bslash \, \aslash)}
\def\ijaP{{\tr}_{+}(\islash\, \jslash\, \aslash \, \Pslash)}
\def\ijPLa{{\tr}_{+}(\islash\, \jslash\, \Pslash_L \, \aslash)}
\def\ijaPL{{\tr}_{+}(\islash\, \jslash\, \aslash \, \Pslash_L)}
\def\ijPLza{{\tr}_{+}(\islash\, \jslash\, \Pslash_{L;z} \, \aslash)}
\def\ijaPLz{{\tr}_{+}(\islash\, \jslash\, \aslash \, \Pslash_{L;z})}
\def\ijPa{{\tr}_{+}(\islash\, \jslash\, \Pslash \, \aslash)}
\def\iaPb{{\tr}_{+}(\islash\, \aslash\, \Pslash \, \bslash)}
\def\ibPa{{\tr}_{+}(\islash\, \bslash\, \Pslash \, \aslash)}
\def\ijPmu{{\tr}_{+}(\islash\, \jslash\, \Pslash \, \mu)}
\def\ibmuP{{\tr}_{+}(\islash\, \bslash\, \mu \, \Pslash)}
\def\ibmua{{\tr}_{+}(\islash\, \bslash\, \mu \, \aslash)}
\def\iamub{{\tr}_{+}(\islash\, \aslash\, \mu \, \bslash)}
\def\jaPb{{\tr}_{+}(\jslash\, \aslash\, \Pslash \, \bslash)}
\def\ijmuP{{\tr}_{+}(\islash\, \jslash\, \mu \, \Pslash)}
\def\ijmum{{\tr}_{+}(\islash\, \jslash\, \mu \, \mslash)}
\def\ijmmu{{\tr}_{+}(\islash\, \jslash\, \mslash \, \mu)}
\def\ijmP{{\tr}_{+}(\islash\, \jslash\, \mslash \, \Pslash)}
\def\iabP{{\tr}_{+}(\islash\, \aslash\, \bslash \, \Pslash)}
\def\ijbP{{\tr}_{+}(\islash\, \jslash\, \bslash \, \Pslash)}
\def\jbPa{{\tr}_{+}(\jslash\, \bslash\, \Pslash \, \aslash)}
\def\ijPb{{\tr}_{+}(\islash\, \jslash\, \Pslash \, \bslash)}
\def\jbmua{{\tr}_{+}(\jslash\, \bslash\, \mu \, \aslash)}

\def\loablt{ {\tr}_{+}(\lslash_1\, \aslash \, \bslash\, \lslash_2)}

\def\ijlolt{{\tr}_{+}(\islash\, \jslash\, \lslash_1 \, \lslash_2)}
\def\ijltlo{{\tr}_{+}(\islash\, \jslash\, \lslash_2 \, \lslash_1)}
\def\ibloa{{\tr}_{+}(\islash\, \bslash\, \lslash_1 \, \aslash)}
\def\jaltb{{\tr}_{+}(\jslash\, \aslash\, \lslash_2 \, \bslash)}
\def\ialtb{{\tr}_{+}(\islash\, \aslash\, \lslash_2 \, \bslash)}
\def\bltloa{{\tr}_{+}(\bslash\, \lslash_2\, \lslash_1 \, \aslash)}
\def\jbloa{{\tr}_{+}(\jslash\, \bslash\, \lslash_1 \, \aslash)}
\def\ibPb{{\tr}_{+}(\islash\, \bslash\, \Pslash \, \bslash)}
\def\ijltb{{\tr}_{+}(\islash\, \jslash\, \lslash_2 \, \bslash)}

\def\ijloa{{\tr}_{+}(\islash\, \jslash\,  \lslash_1 \, \aslash)}
\def\ijblt{{\tr}_{+}(\islash\, \jslash\,  \bslash \, \lslash_2)}

\def\jakb{{\tr}_{+}(\jslash\, \aslash\, \kslash \, \bslash)}
\def\iakb{{\tr}_{+}(\islash\, \aslash\, \kslash \, \bslash)}

\def\tofo{{\tr}_{+}(\onslash\, \thslash\, \twslash \, \foslash)}
\def\foto{{\tr}_{+}(\onslash\, \thslash\, \foslash \, \twslash)}
\def\tofi{{\tr}_{+}(\onslash\, \thslash\, \twslash \, \fislash)}
\def\fito{{\tr}_{+}(\onslash\, \thslash\, \fislash \, \twslash)}

\def\lrangle#1#2{\langle #1\,#2\rangle}

\def\Li{{$\rm Li}_2$}

\let\bm=\bibitem
\let\la=\label

\def\npb#1#2#3{Nucl. Phys. {\bf{B#1}} #3 (#2)}
\def\plb#1#2#3{Phys. Lett. {\bf{#1B}} #3 (#2)}
\def\prl#1#2#3{Phys. Rev. Lett. {\bf{#1}} #3 (#2)}
\def\prd#1#2#3{Phys. Rev. {D \bf{#1}} #3 (#2)}
\def\cmp#1#2#3{Comm. Math. Phys. {\bf{#1}} #3 (#2)}
\def\cqg#1#2#3{Class. Quantum Grav. {\bf{#1}} #3 (#2)}
\def\nppsa#1#2#3{Nucl. Phys. B (Proc. Suppl.) {\bf{#1A}}#3 (#2)}
\def\ap#1#2#3{Ann. of Phys. {\bf{#1}} #3 (#2)}
\def\ijmp#1#2#3{Int. J. Mod. Phys. {\bf{A#1}} #3 (#2)}
\def\rmp#1#2#3{Rev. Mod. Phys. {\bf{#1}} #3 (#2)}
\def\mpla#1#2#3{Mod. Phys. Lett. {\bf A#1} #3 (#2)}
\def\jhep#1#2#3{J. High Energy Phys. {\bf #1} #3 (#2)}
\def\atmp#1#2#3{Adv. Theor. Math. Phys. {\bf #1} #3 (#2)}
\newcommand{\EQ}[1]{\begin{equation} #1 \end{equation}}
\newcommand{\AL}[1]{\begin{subequations}\begin{align} #1 \end{align}\end{subequations}}
\newcommand{\SP}[1]{\begin{equation}\begin{split} #1 \end{split}\end{equation}}
\newcommand{\ALAT}[2]{\begin{subequations}\begin{alignat}{#1} #2 \end{alignat}
                        \end{subequations}}
\def\beqa{\begin{eqnarray}}
\def\eeqa{\end{eqnarray}}
\def\beq{\begin{equation}}
\def\eeq{\end{equation}}
\def\sst{\scriptscriptstyle}
\def\thetabar{\bar\theta}
\def\Tr{{\rm Tr}}
\def\one{\mbox{1 \kern-.59em {\rm l}}}
 \def\Nh{\hat{N}}

\def\a{\alpha}      \def\da{{\dot\alpha}}
\def\b{\beta}       \def\db{{\dot\beta}}
\def\g{\gamma}  \def\G{\Gamma}  \def\cdt{\dot\gamma}
\def\d{\delta}  \def\D{\Delta}  \def\ddt{\dot\delta}
\def\e{\epsilon}        \def\vare{\varepsilon}
\def\f{\phi}    \def\F{\Phi}    \def\vvf{\f}
\def\h{\eta}
\def\k{\kappa}
\def\l{\lambda} \def\L{\Lambda}
\def\m{\mu} \def\n{\nu}
\def\o{\omega}
\def\p{\pi} \def\P{\Pi}
\def\r{\rho}
\def\s{\sigma}  \def\S{\Sigma}
\def\t{\tau}
\def\th{\theta} \def\Th{\Theta} \def\vth{\vartheta}
\def\X{\Xeta}
\def\z{\zeta}

\def\cA{{\cal A}} \def\cB{{\cal B}} \def\cC{{\cal C}}
\def\cD{{\cal D}} \def\cE{{\cal E}} \def\cF{{\cal F}}
\def\cG{{\cal G}} \def\cH{{\cal H}} \def\cI{{\cal I}}
\def\cJ{{\cal J}} \def\cK{{\cal K}} \def\cL{{\cal L}}
\def\cM{{\cal M}} \def\cN{{\cal N}} \def\cO{{\cal O}}
\def\cP{{\cal P}} \def\cQ{{\cal Q}} \def\cR{{\cal R}}
\def\cS{{\cal S}} \def\cT{{\cal T}} \def\cU{{\cal U}}
\def\cV{{\cal V}} \def\cW{{\cal W}} \def\cX{{\cal X}}
\def\cY{{\cal Y}} \def\cZ{{\cal Z}}

\def\ua{\underline{\alpha}}
\def\ub{\underline{\phantom{\alpha}}\!\!\!\beta}
\def\uc{\underline{\phantom{\alpha}}\!\!\!\gamma}
\def\um{\underline{\mu}}
\def\ud{\underline\delta}
\def\ue{\underline\epsilon}
\def\una{\underline a}\def\unA{\underline A}
\def\unb{\underline b}\def\unB{\underline B}
\def\unc{\underline c}\def\unC{\underline C}
\def\und{\underline d}\def\unD{\underline D}
\def\une{\underline e}\def\unE{\underline E}
\def\unf{\underline{\phantom{e}}\!\!\!\! f}\def\unF{\underline F}
\def\unm{\underline m}\def\unM{\underline M}
\def\unn{\underline n}\def\unN{\underline N}
\def\unp{\underline{\phantom{a}}\!\!\! p}\def\unP{\underline P}
\def\unq{\underline{\phantom{a}}\!\!\! q}
\def\unQ{\underline{\phantom{A}}\!\!\!\! Q}
\def\unH{\underline{H}}

\def\As {{A \hspace{-6.4pt} \slash}\;}
\def\bs {{b \hspace{-6.4pt} \slash}\;}
\def\Ds {{D \hspace{-6.4pt} \slash}\;}
\def\ds {{\del \hspace{-6.4pt} \slash}\;}
\def\ss {{\s \hspace{-6.4pt} \slash}\;}
\def\ks {{ k \hspace{-6.4pt} \slash}\;}
\def\ps {{p \hspace{-6.4pt} \slash}\;}
\def\pas {{{p_1} \hspace{-6.4pt} \slash}\;}
\def\pbs {{{p_2} \hspace{-6.4pt} \slash}\;}
\def\Ps {{P \hspace{-6.4pt} \slash}\;}
\def\Qs {{Q \hspace{-6.4pt} \slash}\;}

\def\Fh{\hat{F}}
\def\Vh{\hat{V}}
\def\Xh{\hat{X}}
\def\ah{\hat{a}}
\def\xh{\hat{x}}
\def\yh{\hat{y}}
\def\ph{\hat{p}}
\def\xih{\hat{\xi}}
\def\psit{\tilde{\psi}}
\def\Psit{\tilde{\Psi}}
\def\tht{\tilde{\th}}
\def\lt{\tilde{\lambda}}
\def\llt{\tilde{l}}
\def\At{\tilde{A}}
\def\Qt{\tilde{Q}}
\def\Rt{\tilde{R}}
\def\Nt{\tilde{N}}

\def\at{\tilde{a}}
\def\st{\tilde{s}}
\def\ft{\tilde{f}}
\def\pt{\tilde{p}}
\def\qt{\tilde{q}}
\def\vt{\tilde{v}}
\def\nt{\tilde{n}}

\def\delb{\bar{\partial}}
\def\bz{\bar{z}}
\def\bD{\bar{D}}
\def\bB{\bar{B}}

\def\bk{{\bf k}}
\def\bl{{\bf l}}
\def\bp{{\bf p}}
\def\bq{{\bf q}}
\def\br{{\bf r}}
\def\bx{{\bf x}}
\def\by{{\bf y}}
\def\bR{{\bf R}}
\def\bV{{\bf V}}

\def\d{\delta}\def\D{\Delta}\def\ddt{\dot\delta}
\def\pa{\partial} \def\del{\partial}
\def\xx{\times}
\def\uno{\mbox{1 \kern-.59em {\rm l}}}
\def\trp{^{\top}}
\def\inv{^{-1}}
\def\dag{{^{\dagger}}}
\def\pr{^{\prime}}
\def\lan{\langle}
\def\ran{\rangle}
\def\rar{\rightarrow}
\def\lar{\leftarrow}
\def\lrar{\leftrightarrow}
\newcommand{\0}{\,\!}      
\def\one{1\!\!1\,\,}
\def\im{\imath}
\def\jm{\jmath}
\newcommand{\tr}{\mbox{tr}}
\newcommand{\slsh}[1]{/ \!\!\!\! #1}
\def\vac{|0\rangle}
\def\lvac{\langle 0|}
\def\hlf{\frac{1}{2}}
\def\ove#1{\frac{1}{#1}}
\def\Box{\square}
\def\ZZ{\mathbb{Z}}
\def\CC#1{({\bf #1})}
\def\bcomment#1{}
\def\bfhat#1{{\bf \hat{#1}}}
\def\VEV#1{\left\langle #1\right\rangle}
\newcommand{\ex}[1]{{\rm e}^{#1}} \def\ii{{\rm i}}
\def\rr{{\rm r}} \def\rs{{\rm s}}\def\rv{{\rm v}}
\def\ri{{\rm i}}\def\rj{{\rm j}}
\newcommand{\lrbrk}[1]{\left(#1\right)}
\newcommand{\sfrac}[2]{{\textstyle\frac{#1}{#2}}}

\def\Li{{\rm Li}_2}

\def\Li2{{\rm Li}_2}

\def\intD{\int\!{d^DL \over (2\pi)^D}}

\font\mybb=msbm10 at 12pt
\def\bb#1{\hbox{\mybb#1}}

\font\myBB=msbm10 at 18pt
\def\BB#1{\hbox{\myBB#1}}

\begin{document}

\begin{flushright}
hep-th/0612007\\
QMUL-PH-06-12
\end{flushright}

\vspace{20pt}

\begin{center}

{\Large \bf  Amplitudes in Pure Yang-Mills }\\
\vspace{0.3cm}
{\Large \bf and MHV Diagrams}
\vspace{33pt}

{\bf {\mbox{Andreas  Brandhuber,  Bill Spence and Gabriele  Travaglini}}}%
\footnote{{\sffamily \{\tt a.brandhuber, w.j.spence, g.travaglini\}@qmul.ac.uk }}

{\em Centre for Research in String Theory \\ Department of Physics\\
Queen Mary, University of
London\\
Mile End Road, London, E1 4NS\\
United Kingdom}
\vspace{40pt}

{\bf Abstract}

\end{center}

\noindent

We  show how to calculate the one-loop scattering amplitude
with all gluons of negative
helicity in non-supersymmetric Yang-Mills theory using MHV diagrams.
We argue that the
amplitude with all positive helicity gluons arises from a Jacobian
which occurs when one performs a B\"{a}cklund-type holomorphic change of
variables in the lightcone Yang-Mills Lagrangian. This also results in
contributions to scattering amplitudes from violations of the equivalence
theorem. Furthermore, we discuss how the
one-loop amplitudes with a single positive or negative helicity gluon arise
in this formalism.
Perturbation theory in the new variables leads to a hybrid of MHV diagrams and
lightcone Yang-Mills theory.
\vspace{0.5cm}

\setcounter{page}{0}
\thispagestyle{empty}
\newpage


\setcounter{footnote}{0}

\section{Introduction}

Witten's twistor string theory \cite{witten} has prompted
 many new developments, particularly in the study of perturbative
 gauge theories and gravity (see \cite{Cachazo:2005ga} and references therein).
 One might group these into the categories of new
 unitarity-based methods, and what we will call the MHV approach (see
 \cite{Brandhuber:2006vh} for a review), which will
 be most relevant to this paper.
The MHV approach originates in the work of \cite{csw}, who showed that tree amplitudes in
Yang-Mills theories could be derived by sewing together MHV vertices, suitably continued
off shell. This approach was subsequently shown to work at one-loop level, with the derivation
of the complete MHV amplitudes for $\cN=4,1$ super Yang-Mills \cite{bst, quig,bbst1}, and the
cut-constructible part of the MHV amplitudes for pure gauge theory \cite{bbst2}.
Proofs of the method at  tree level were then presented in
\cite{rec,ris}.
More recently, in \cite{ftt} strong evidence was presented in favour of
the correctness of the MHV diagram method
at one loop, when applied to supersymmetric theories. In particular,
it was shown in \cite{ftt} that the MHV diagram
calculation of a generic one-loop amplitude
in supersymmetric theories correctly reproduces all discontinuities,
as well as collinear and soft limits.

Whilst this technique gives the right answers in the cases studied so far,
a complete proof at the quantum level is still missing.
A further problem, pertinent for the  practical
use of this method, is that MHV diagrams
have so far only yielded the cut-constructible part of amplitudes,
missing rational terms (and furthermore certain amplitudes in pure Yang-Mills and QCD 
are entirely rational).%
\footnote{For recent progress in evaluating such rational terms, see
\cite{b1,b2,b3,z1,z2,z3,Anastasiou:2006jv,pm}.}

Recently, progress has been made in systematising the MHV approach. Study of the lightcone
gauge-fixed Yang-Mills Lagrangian has shown that if one uses a certain change of variables,
then one may write the Lagrangian as a kinetic term plus an infinite sequence of MHV
interaction terms \cite{Gorsky:2005sf, Mansfield, Ettle:2006bw}. As far as the classical level
is concerned, this provides a Lagrangian description of the MHV diagrams approach to tree-level
amplitudes. If one were able to formulate the quantum theory using similar ideas, then one
would have a prescription for an alternative perturbation theory for gauge theories, based on
MHV diagrams. This would carry the great practical advantage of being much more suited to the
calculation of amplitudes than the usual Feynman rules, which
become completely unwieldy when
dealing with more than a small number of particles. Given the close relationship of the twistor
localisation of amplitudes with MHV diagrams, this may also lead to a more direct twistor space
formulation of gauge theories than we currently have available.

How might one go about formulating a quantum version of the analysis presented in
 \cite{Gorsky:2005sf, Mansfield, Ettle:2006bw}? One of
the obvious omissions is that of understanding how one derives the
 purely rational
amplitudes, for example the all-plus helicity amplitudes in pure Yang-Mills.
These cannot be
generated from MHV diagrams, which can only give rise to amplitudes with more than
one gluon with negative helicity. 
There appears to be no way to obtain these using the ideas of
\cite{Gorsky:2005sf, Mansfield, Ettle:2006bw}, since they do
not appear in the Lagrangian and cannot be generated from it, nor are there any
measure-related terms, since the change of variables used is canonical.

However, we will first show how the amplitudes with all {\it negative} helicity particles
in pure Yang-Mills arise from MHV diagrams, which has not been done so far. 
This involves the use of three-point vertices which we
observe are precisely the same as the lightcone gauge vertices. This
implies that the all-minus amplitudes calculated with MHV diagrams will
give the same results as those found from Feynman rules in the lightcone gauge.
We also show this explicitly for the cases of three and four external particles. This involves
the derivation of the three-particle all-minus vertex with one leg off shell. These calculations 
confirm the above argument 
that all of the one-loop all-minus amplitudes are generated from MHV diagrams in the
same manner. 

We then consider a certain holomorphic change of variables in lightcone gauge theory (the
redefinition of \cite{Mansfield} involves fields of  both helicity). This change of variables
generates a Jacobian factor in the measure, and
we argue that one-loop diagrams leading to
all-plus amplitudes are generated from this.
We then consider the amplitudes which have a single negative helicity, or a single
positive helicity particle, and we discuss how these arise using the new variables.
Contributions coming from violations of the equivalence theorem play an important r\^{o}le here.
Finally we discuss the perturbation theory which arises in this formulation of Yang-Mills
theory and which could be called a hybrid MHV/lightcone perturbation theory.

The outline of the paper is as follows. In section 2 we lay out the strategy for calculating the
all-minus helicity one-loop amplitude and recollect some facts about lightcone perturbation
theory and MHV diagrams. Then we proceed to calculate the three-minus one-loop vertex with
one leg off-shell in section 3. This result is used in section 4 to calculate the four-point all-minus
one-loop amplitude from MHV vertices. This calculation gives the correct answer to all orders in
the dimensional regularisation parameter $\epsilon$. In section 5 we consider a holomorphic
field redefinition which is non-canonical (unlike Mansfield's transformation) and leads to a 
non-trivial Jacobian in the path integral. We show that this Jacobian incorporates precisely the
missing vertices to generate the all-plus one-loop amplitudes. 
Furthermore, we discuss one-loop amplitudes with all but one gluon of the same helicity. 
Although these two types of amplitudes are
related by complex conjugation they have quite different origins in our formalism, and we find
that the violation of the equivalence theorem  by the holomorphic field redefinition plays an important r\^{o}le.
This suggests a novel kind of perturbation theory for Yang-Mills which combines  
MHV-type vertices,  vertices from the Jacobian (which account for the all-plus amplitude), 
and correction terms from
the violation of the equivalence theorem. 
Conclusions can be found in section 6.

\section{The all-minus amplitudes}

In this section we will discuss the derivation of the all-minus helicity one-loop amplitudes
in pure Yang-Mills theory using MHV vertices. We first find the three-particle
amplitude, using fundamental three-point scalar-gluon vertices, which we will
then continue off shell to obtain the relevant MHV vertices.
We then show that the MHV diagram calculation
is the same as that using the lightcone gauge-fixed Lagrangian, and hence that
all the all-minus amplitudes calculated with MHV diagrams yield the same
results as the lightcone approach. In sections 3 and 4 we will check this against
explicit one-loop calculations. Specifically, in section 3 we work out a three-point all-minus
vertex, and in section 4 we use this result to derive the four-point all-minus amplitude.

\subsection{Coupling to a scalar}
In the following we will make use of the supersymmetric decomposition
of one-loop amplitudes of gluons
in pure Yang-Mills.
If $\cA_{\rm g}$ is a certain gluon scattering amplitude
with gluons running in the loop, one can decompose it as
\beq
\label{susydec}
\cA_{\rm g} \ = \ (\cA_{\rm g} \, + \, 4 \cA_{\rm f} \, + \,
3 \cA_{\rm s}) \  - \
4( \cA_{\rm f}  + \cA_{\rm s}) \ + \  \cA_{\rm s}
\ .
\eeq
Here $\cA_{\rm f}$ ($\cA_{\rm s}$) is the amplitude with the same
external particles as   $\cA_{\rm g}$ but with a Weyl fermion
(complex scalar) in the adjoint of the gauge group running
in the loop.
The first two terms on the right hand side of
\eqref{susydec} are  contributions
coming from an $\cN \! = \! 4$ multiplet and
(minus four times) a chiral $\cN \! = \! 1$ multiplet,
respectively.
The last term in \eqref{susydec}, $\cA_{\rm s}$,
is the contribution arising from a scalar running in the loop.
The key point here is that the calculation
of this term is simpler than that
of the original amplitude $\cA_{\rm g}$ with a gluon running in the loop,
since for a scalar we do not have to worry about
$D=4-2 \epsilon$ dimensional polarisation vectors.
Since the all-minus amplitude is zero to all orders of perturbation theory
in any supersymmetric theory, we have to calculate only the
last contribution. For this reason we will now focus on the coupling to
a complex scalar field.

We start from
\beq
\cL_{\rm s} := \overline{(D_\m \phi)} (D^\m \phi)
\ ,
\eeq
where $D_\m$ is the adjoint derivative. In the gauge $A^- = 0$
the interaction terms are%
\footnote{A brief review of lightcone quantisation of Yang-Mills theory with a
summary of our conventions
is contained in appendix \ref{lcym}.}
\beqa
\cL_{3, {\rm s}} &=& i A^+ \Big( [ \phi, \del^- \bar{\phi}] + [ \bar{\phi} , \del^-  \phi]\Big)
- i \Big( \del^{\hat\m} \bar{\phi} [ A^{\hat\m} , \phi] + \del^{\hat\m}\phi [A^{\hat\mu} , \bar{\phi}] \Big)
\ ,
\nonumber \\
\cL_{4, {\rm s}} &= &- (\phi \bar\phi + \bar\phi \phi ) A^{\hat\m}  A^{\hat\m} + 2 \bar{\phi} A^{\hat\m} \phi A^{\hat\m}
\ .
\eeqa
Integrating out $A^+$ is equivalent to replacing $A^+$ by
\beq
A^+ = (\del^-)^2 ( \del^{\hat\m} \del^- A^{\hat{\m}} + i [    A^{\hat\m}, \del^- A^{\hat\m}] +
i [ \phi , \del^- \bar\phi] + i [ \bar\phi, \del^- \phi] )
\ .
\eeq
The lightcone Lagrangian for the coupling of the scalar to the gauge field
is given by
\beq
\cL_{\rm s} = \cL_{\rm s}^{(2)} + \cL_{\rm s}^{(3)} + \cL_{\rm s}^{(4)}
\ .
\eeq
Here
$ \cL_{\rm s}^{(2)} = (-1/2) \phi^a \Box \phi^a$ is the kinetic term, and
\beqa
\cL_{\rm s}^{(3)}  &=&
i \left( [\phi^a , \del^- \phi^a] (\del^-)^{-1} \del^{\hat\m} A^{\hat\m} -
[ \phi^a , \del^{\hat\m} \phi^a] A^{\hat\m} \right)
\ ,
\nonumber \\
\cL_{\rm s}^{(4)}  &=&
-{1\over 2} [\phi^a , \del^- \phi^a](\del^-)^{-2}[\phi^b, \del^- \phi^b]
- [\phi^a , \del^- \phi^a](\del^-)^{-2} [ A^{\hat\m} , \del^- A^{\hat\m}]
\cr
&&
- \phi^a \phi^a A^{\hat\m}A^{\hat\m} + \phi^a A^{\hat\m}\phi^a A^{\hat\m}
\ ,
\eeqa
where we have introduced the  notation
$\phi^a \cdots \phi^a = \phi \cdots \bar\phi + \bar\phi \cdots \phi$.

We use the four-dimensional helicity scheme, where the momenta of external (gluon) particles
are kept in four dimensions, and those of the internal loop particles are in $D = 4-2\e$ dimensions.
We will consider only scalars propagating in the loop, hence we can restrict our attention
to four-dimensional gauge fields. Then
one finds that $\cL_{\rm s}^{(3)} $ can be re-written as
\beq
\cL_{\rm s}^{(3)}   = \cL_{\phi \bar\phi A_z} + \cL_{\phi \bar\phi A_{\bz}}
\ ,
\eeq
where
\beqa
\cL_{\phi \bar\phi A_z} &=& i \Big( [ \phi^a , \del^- \phi^a ] (\del^-)^{-1} \del_{\bz} -
[ \phi^a , \del_{\bz}  \phi^a ] \Big) A_z
\ , 
\\ \nonumber
\cL_{\phi \bar\phi A_{\bz}} &=& i \Big( [ \phi^a , \del^- \phi^a ] (\del^-)^{-1} \del_{z} -
[ \phi^a , \del_{z}  \phi^a ] \Big) A_{\bz}
\ .
\eeqa
$A_z$ and $A_{\bz}$ are defined in \eqref{azs} and create gluons in states of  definite 
helicity.

\subsection{Three-point vertices}
In order to calculate the all-plus or the all-minus amplitude from the lightcone
Lagrangian, we only need to work out the three-point interaction between scalars and
gluons. Any other vertex cannot contribute to an amplitude
with all same-helicity gluons. More precisely, $\cL_{-++}$ ($\cL_{--+}$)
and $\cL_{\phi \bar\phi A_z}$ ($\cL_{\phi \bar\phi A_{\bz}}$) are the only terms which
can contribute to an all-plus (all-minus) gluon amplitude with complex scalars
running in the loop. Indeed, it is well known that the all-plus (all-minus) amplitude
in pure Yang-Mills can equivalently be computed using the self-dual (anti-self-dual)
truncation of Yang-Mills.

After a short calculation which makes use of 
\beq
\label{app111}
{ [ \eta | l |  k\ran \over [\eta k] }  = \sqrt{2} {l_+k_{\bar z} - l_{\bar z}k_+ \over k_+ }
\ ,
\eeq
where 
\beq
\eta^a = \left (\ \begin{matrix} { 0} \\ {1 } \end{matrix}\ \right) \ ,
\eeq
we find that the explicit form of the three-point vertices is given by
\beqa
\label{3pv1}
v^{(3)}_{\bar\phi   A_z \phi} (L_1, k, L_2)&  =& {\lan \eta | L_1| k] \over \lan \eta k \ran}
\ ,
 \\
\label{3pv2}
v^{(3)}_{\bar\phi  A_{\bz} \phi} (L_1, k, L_2)  & = & {[ \eta | L_1| k\ran \over [ \eta k ] }
\ .
\eeqa

We now make some observations:

{\bf 1.} Firstly, in the expressions \eqref{3pv1},  \eqref{3pv2}, one can drop the $-2\e$-dimensional part of
$L_1^{(D)}$, as it is contracted between four-dimensional spinors.

{\bf 2.}
As observed above, we will consider only scalars running in the loop.
If $L_D$ is the $(4-2\e)$-dimensional momentum of the
massless scalar, we will decompose it into a
four-dimensional component $L$ and a $ -2\e$-dimensional
component $L_{(-2 \e)}$, $L _D:= L + L_{(-2\e)}$.
Then  $L^2_D := L^2 +  L_{(-2\e)}^2 = L^2 - \mu^2$,
where $L_{(-2\e)}^2 := - \mu^2$,  and the four-dimensional
and $-2\e$-dimensional subspaces are taken to be orthogonal.

{\bf 3.}
In general we can decompose any four-momentum $L $ as
\beq
\label{off}
L :=  l + z \eta
 \ ,
 \eeq
where $l^2=0$. It follows that in equations \eqref{3pv1} and  \eqref{3pv2}  we can replace
(the four-dimensional part of) $L_1$ by its on-shell, lightcone truncation.
Hence, we can rewrite
\beqa
\label{3pvnew1}
v^{(3)}_{\bar\phi A_z \phi} (l_1, k, l_2)&  =& {\lan \eta | l_1|  k] \over \lan \eta k \ran}
\ ,
\\
\label{3pvnew2}
v^{(3)}_{\bar\phi  A_{\bz} \phi} (l_1, k, l_2)  & = & {[ \eta | l_1| k\ran \over [ \eta k ] }
\ ,
\eeqa
with $L_1 = l_1 + z \eta$.

{\bf 4.}
The final comment we would like to make is that \eqref{3pvnew2}  (\eqref{3pvnew1})
is nothing but the MHV (anti-MHV) three-point
vertex for two scalars and a gluon of negative (positive) helicity,
continued off-shell using the prescription introduced by Cachazo, Svr\v{c}ek and
Witten in \cite{csw},
if the arbitrary null vector $\eta$ introduced in the MHV rules is chosen to be the same as that
introduced to pick the lightcone gauge \eqref{lc}.
In order to see this, we notice that the MHV scalar-scalar-gluon vertex is
\beq
\label{v3mhv}
V_3 = { \lan l_1 k \ran \lan l_2 k \ran \over \lan l_1 l_2 \ran}
\ .
\eeq
Here, to an off-shell vector $L$ one associates a null vector $l_{\a \da} = l_{\a} \llt_{\da}$ using
\eqref{off}, where $\eta$ is at this point an arbitrary null vector.
Specifically,
\beq
\label{1}
l_\a \  = \  {L_{\a \da} \tilde{\eta}^{\da}
\over [ \llt \, \tilde{\eta}]
}
\ ,
\qquad
 \llt_{\da} \ = \ {\eta^\a L_{\a \da} \over \lan l \, \eta
\ran} \ .
 \eeq
This is the  CSW prescription \cite{csw} for
determining the spinor variables  $l_{\a }$ and $\llt_{\da}$ associated with
any off-shell (i.e.~non-null) four-vector $L$.%
\footnote{The denominators on the right hand sides of the two expressions in
\eqref{1} are not present in the CSW off-shell continuation but  are in fact irrelevant,
since the expressions we will be dealing with are homogeneous in the spinor
variables $\eta$.}
By selecting  $\eta$ to be the same null vector defining
the lightcone gauge \eqref{lc},  the lightcone truncation of a
generic vector $L$ coincides with the null momentum $l$ appearing
in the MHV rules. Taking this into account, it is immediate to show that
\eqref{3pvnew2} and \eqref{v3mhv} are identical.

This conclusion is not surprising -- as mentioned above, in a very interesting paper
\cite{Mansfield}, Mansfield was able to construct a
MHV Lagrangian via a canonical change of variables of the
lightcone Yang-Mills Lagrangian \eqref{lymlc}. This nonlocal change of variables
has the effect of removing the anti-MHV interaction $\cL_{++-}$ from the theory,
at the expense of introducing an infinite number of MHV vertices
(see also \cite{Gorsky:2005sf,Ettle:2006bw}).
Schematically, the structure of this change of variables is,
\beqa
\label{mansfield}
A_z &=& B_z  \big(1 + \cO ( B_z ) \big)
\ ,
\nonumber \\
A_{\bz} &=& B_{\bz} \big( 1 + \cO ( B_z )  \big)
\ ,
\eeqa
and is engineered in such a way that the Lagrangian
of self-dual Yang-Mills is mapped to that of a free theory
\cite{Mansfield},
\beq
(\cL_{-+} + \cL_{++-})(A_z, A_{\bz}) = \cL_{-+} (B_z , B_{\bz})
\ .
\eeq
It is clear that, upon substituting \eqref{mansfield} into
$\cL_{--+}$ and $\cL_{--++}$,  the new three-point $--+$ vertex
(i.e.~the three-point MHV vertex)
is identical to the corresponding vertex in the original lightcone Lagrangian.

This simple observation, together with the previous remark that
only three-point MHV vertices can contribute to the all-minus amplitude,
is sufficient to guarantee that the all-minus amplitude  derived using
MHV rules will be correct, to all orders in $\e$.
This is because the MHV diagram calculation
in this specific case is nothing but the lightcone Yang-Mills calculation.

It will prove illuminating for later sections
to present the explicit calculation of the simplest such amplitudes.
This will also explicitly show how these amplitudes arise from an
$\e \times 1/ \e$ cancellation in dimensional regularisation.
Specifically, in the following we will discuss two calculations:
that of a four-point all-minus amplitude, and, in preparation to this, that of
a three-point one-loop {\it vertex}, with one leg continued off shell.


\section{The one-loop three-minus vertex}
In this section we will derive an expression for the one-loop amplitude
with three gluons  of negative helicity, where one of the external legs is continued off shell.
This vertex, depicted in figure 1, will enter the calculation, to be discussed in the next section, of the
four-point all-minus scattering amplitude.

\begin{figure}[ht]
\label{threevertex}
\begin{center}
\scalebox{0.3}{\includegraphics{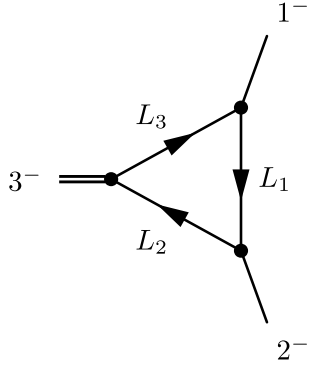}}
\end{center}
\caption{\it
The MHV-diagram for the all-minus three-gluon vertex with one leg off shell
(drawn with a double line).
}
\end{figure}

Using the three-point vertex \eqref{3pv2}, the three-minus one-loop vertex is proportional
to the following integral:
\beq
\label{I}
\cT := \int\!{d^DL \over (2\pi)^D} {[\eta | L_1 |1\ran [\eta | L_2 |2\ran [\eta | L_3 |3\ran \over [\eta 1] [\eta 2] [\eta 3]}
\cdot {1 \over L_{1,D}^2 L_{2,D}^2 L_{3,D}^2} \ .
\eeq
We will keep leg 3 off shell, so that
 the spinors associated to that leg are defined
via the lightcone off-shell continuation \eqref{off}.
Notice also that we keep the propagators $D$-dimensional.

Let us focus on the quantity%
\footnote{As remarked earlier, we could equivalently write $l_1$ and $l_2$
in place of $L_1$ and $L_2$ in  \eqref{I}. We choose to keep
$L_1$ and $L_2$ in order to simplify the $D$-dimensional Passarino-Veltman
reductions to be performed later.}
\beq
\Sigma := {[\eta | L_1 |1\ran [\eta | L_2 |2\ran  \over [\eta 1] [\eta 2] }
\ ,
\eeq
which appears in \eqref{I}.
As in  \cite{Bern:1995db,bmst}, we write
\beq
{   [\eta\vert L_1\vert 1\ran [\eta\vert L_2\vert 2\ran
           \over
              [\eta 1][\eta 2]}   =
{ [\eta\vert \hat L_1 \hat k_1  \hat k_2 \hat L_1 \vert \eta ]
             \over
 [\eta 1] [ \eta 2] [12]  }
 \ ,
\eeq
where we have used $L_2 = L_1-k_2$   and $\hat k_2^2 = 0$.
Note that $\hat L_1 \hat 1 = - \hat 1 \hat L_1 + 2 (k_1\cdot L_1)$, 
$\hat 2\hat L_1  = -  \hat L_1 \hat 2 + 2 (k_2\cdot L_1)$ and
$L_2^2 = L_1^2 - 2(L_1\cdot k_2)$. Hence, we find that
$2(l_1\cdot k_2) = L_1^2-L_2^2$ and 
$2(l_1\cdot k_1) = L_4^2-L_1^2$. Using these relations we arrive at
\beq
\label{ff}
\Sigma =
- L_1^2 {[12] \over \lan1 2\ran} + (L_3^2 - L_1^2) {[\eta | L_1 |2 \ran \over [\eta 1]}
+ (L_2^2 - L_1^2) {[\eta | L_1 |1 \ran \over [\eta 2]}
\ .
\eeq
We remark that the
spinor algebra from the MHV vertices is four-dimensional, but the propagators
are $D$-dimensional.%
\footnote{We would like to point out  that, in the integration measure
derived in \cite{bst}, the propagators are also written explicitly in $D$-dimensions.
This is why the phase-space integrals appearing in that paper are also
$D$-dimensional.}
Note that
\beq
\label{crucial}
{L^2 \over L_{D}^2} = {L^2 - \mu^2 + \mu^2 \over L_{D}^2} = 1 +
{\mu^2 \over L_{D}^2}
\ ,
\eeq
i.e.~incomplete cancellations of propagators have to be properly
taken into account. In supersymmetric theories this is not needed, at
least at one loop,
due to four-dimensional cut-constructibility
\cite{Bern:1994cg}. However, for the non-supersymmetric amplitude
we are focussing on now, it is crucial to keep track of terms such as
$\mu^2 / L_D^2$ in
\eqref{crucial}. It is precisely these terms that will give rise to the
 correct amplitudes;
in other words, the result of a calculation for the (finite) all-minus
amplitude fully performed in
four dimensions would na\"{i}vely be zero.

Such an incomplete cancellation happens for the first term in \eqref{ff}, where
we use $L_1^2 = L_{1, D}^2 + \mu^2$. On the other hand, we can rewrite
 $L_2^2-L_1^2=  L_{2,D}^2-L_{1,D}^2$, as the $\mu^2$ terms cancel,
as well as   $L_{3}^2-L_{1}^2= L_{3,D}^2-L_{1,D}^2$.
Therefore, we find
\beq
\label{ww}
\Sigma = -\mu^2  {\lan 1 2 \ran \over [12]} + {X \over [12]}
\ ,
\eeq
where
\beq
X:= L_{1,D}^2 {[\eta | L_3 (k_1 + k_2) | \eta] \over [\eta 1] [\eta 2]} + L_{3,D}^2
{[\eta | L_1 | 2\ran \over [\eta 1] } +  L_{2,D}^2
{[\eta | L_1 | 1\ran \over [\eta 2] }
\ .
\eeq
Notice the presence of the $\mu^2$ term in \eqref{ww}.

Using \eqref{ww}, we can  write $\cT$ in \eqref{I} as
\beq
\cT = \cT_A + \cT_B
\ ,
\eeq
where
\beq
\label{IA}
\cT_A := \int\!{d^DL \over (2\pi)^D}  \Big( -\mu^2 {\lan 12 \ran \over [12]}
{[\eta | L_3 |3\ran \over  [\eta 3]} \Big)
\cdot {1 \over L_{1,D}^2 L_{2,D}^2 L_{3,D}^2} \ ,
\eeq
and
\beq
\label{IB}
\cT_B := \int\!{d^D L \over (2\pi)^D}  \Big( { X \over [12]}
{[\eta | L_3 |3\ran \over  [\eta 3]} \Big)
\cdot {1 \over L_{1,D}^2 L_{2,D}^2 L_{3,D}^2}
\ .
\eeq
We first focus on $\cT_A$, and recast it as
\beq
\cT_A = 
{i\over (4 \pi)^{2-\e}}\,
 I [ \mu^2; L_{3 \nu}]
{\lan 12 \ran \over [12] } {[\eta | \nu | 3 \ran \over [\eta 3] }
 \ ,
\eeq
where $I [ \mu^2; L_{3 \nu}]$ is a one-mass vector triangle integral%
\footnote{We summarise our notation and results for integrals in Appendix C.}.
It is given by
\beq
I [ \mu^2; L_{3 \nu}] = \Big(- J_3 (k_3^2) + 2 {J_2 (k_3^2) \over k_3^2} \Big) k_{2\nu} +
\Big(- {J_2 (k_3^2) \over k_3^2}  - J_3 (k_3^2)  \Big)k_{3\nu}
\ .
\eeq
Substituting this into \eqref{IA}, we find
\beq
\cT_A = {i\over (4 \pi)^{2-\e}}\, {\lan 12 \ran \over [12] } {[\eta | 2 | 3 \ran \over [\eta 3] }
\Big(- J_3 (k_3^2) + 2 {J_2 (k_3^2) \over k_3^2} \Big)
\ .
\eeq
Furthermore, it is easy to see that  $\cT_B=0$. Indeed, $\cT_B$
is a sum of three two-tensor bubble integrals, each of which separately vanishes due
to the spinor contractions.

Therefore, the final result for the three-point vertex is
\beq
\label{final3}
V_3 (\e) \ = \  {i\over (4 \pi)^{2-\e}}\, 
  {\lan 12 \ran \over [12] } {[\eta | 2 | 3 \ran \over [\eta 3] }
\Big(- J_3 (k_3^2) + 2 {J_2 (k_3^2) \over k_3^2} \Big)
\ ,
\eeq
where $k_3$ is the off-shell leg, $k_3^2 \neq 0$.
It is illuminating to take the $\e\to 0$ limit of \eqref{final3}.
Since $J_3 (k_3^2) \to -1/2$, $J_2 \to -k_3^2 /6$ in this limit, we get
\beq
\label{final3fin}
V_3 =   {i \over 96 \pi^2}{\lan 12 \ran \over [12] } {[\eta | 2 | 3 \ran \over [\eta 3] }
\ .
\eeq

\subsection{The one-loop splitting amplitude from the one-loop vertex}
As a quick application of the result \eqref{final3fin}
for the four-dimensional limit of the
three-point all-minus vertex we can re-derive the
one-loop splitting amplitude $S_{-} (a^-, b^-)$ 
with one real scalar running in the loop
 \cite{Bern:zx,bdkrec}.

To proceed we consider our previous result
\eqref{final3fin},
\beq
V_3 =   {i \over 96\pi^2}{\lan a b \ran \over [ab] } {[\eta | b | k \ran \over [\eta k] }
\ ,
\eeq
where
$k = -(k_a + k_b)$. In the limit where $k_a$ and $k_b$ become collinear we set, as usual,
$k_a \to zk$, $k_b \to (1-z)k$, and take $k^2 \to 0$. Then we rewrite
\beq
{[\eta | b | k \ran \over [\eta k ] }  =  {\lan a k\ran \lan  kb \ran \over \lan a b \ran} =
\lan ab \ran  { [a \eta ] [b \eta] \over [k \eta]^2}
\ ,
\eeq
where in the first equality we multiplied and divided by $\lan kb \ran$, and in the second by $[k \eta]^2$.
The derivation of the splitting amplitude is similar to that of the tree-level splitting amplitude
for the helicity configuration $ -- \to - $ considered in \cite{csw}, where the splitting amplitude
is obtained by multiplying a vertex by a propagator $1/ k^2$.
Doing this, we obtain
\beq
S_{-}(a^- , b^-) = {i \over 96\pi^2}{\lan a b \ran \over \, \, [ab]^2 } { [a \eta ] [b \eta] \over [k \eta]^2}
\ .
\eeq
Finally replacing $k_a \to zk$, $k_b \to (1-z)k$ we get
\beq
S_{-} (a^-, b^-)(z)  \ = \  {i \over 96 \pi^2} \sqrt{z(1-z)} {\lan ab \ran \over \, \, [ab]^2}
\ = \ 
 {i\over 96 \pi^2} \, \sqrt{z(1-z)} {\lan ab \ran \over \, \, [ab]^2}
\ ,
\eeq
in agreement with the result  of \cite{Bern:zx,bdkrec}.


\section{The four-point all-minus amplitude}
In order to calculate the four-point all-minus
amplitude, we will have to sum diagrams with three different
topologies, namely a box diagram, diagrams containing a
three-point all-minus vertex, and finally bubble-like diagrams.
The box diagram is depicted in figure 2, while the other two
topologies can be found in figure 3.
We will see that the box diagrams contains the correct result for the amplitude,
plus additional terms which will be cancelled by the diagrams containing a
three-minus vertex. The bubble-like diagrams will be seen to vanish.
In the following we present a detailed analysis of these contributions.

\subsection{The box diagram}

\begin{figure}[ht]
\begin{center}
\scalebox{0.3}{\includegraphics{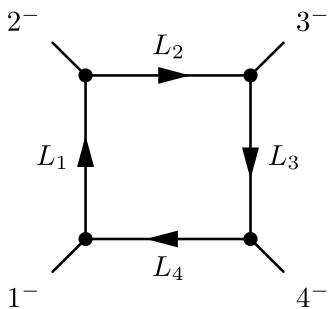}}
\end{center}
\caption{\it
The  box MHV-diagram contributing to the all-minus four-gluon
amplitude. A complex scalar runs in the loop, and one has to sum over the 
two possible internal helicity assignments. This has the effect of doubling  
the result of a single internal helicity assignment.
}
\label{boxdiagram}
\end{figure}

This diagram is depicted in figure \ref{boxdiagram}, and its expression is
\beq
\label{boxini}
\cB =  \intD {   [\eta\vert L_1\vert 1\ran [\eta\vert L_2\vert 2\ran
                      [\eta\vert L_3\vert 3\ran [\eta\vert L_4\vert 4\ran
           \over
              [\eta 1][\eta 2][\eta 3][\eta 4] } { 1\over L_{1,D}^2 L_{2,D}^2 L_{3,D}^{2} L_{4,D}^{2}} \ .
\eeq
Manipulations  similar to those which led to \eqref{ww}
allow us to rewrite

\beq
{   [\eta\vert L_1\vert 1\ran [\eta\vert L_2\vert 2\ran
           \over
              [\eta 1][\eta 2]}   =
 -\mu^2 { \lan 12\ran\over [12] }  + {X\over [12] }
 \ ,
\eeq
where
\beq
    X = L_{1,D}^2  { [\eta \vert L_4(1+2)\vert\eta ] \over  [\eta 1] [ \eta 2] }
 + L_{4,D}^2   { [ \eta\vert L_1\vert 2\ran \over [\eta 1] }
  + L_{2,D}^2 { [ \eta\vert L_1\vert 1\ran \over [\eta 2] }
  \ ,
  \eeq
  and
  \beq
{   [\eta\vert L_3\vert 3\ran [\eta\vert L_4\vert 4\ran
           \over
              [\eta 3][\eta 4]}   =
 -\mu^2 { \lan 34\ran\over [34] }  + {Y\over [34] }
 \ ,
\eeq
where
\beq
    Y = L_{3,D}^2  { [\eta \vert L_2(3+4)\vert\eta ] \over  [\eta 3] [ \eta 4] }
 + L_{2,D}^2   { [ \eta\vert L_3\vert 4\ran \over [\eta 3] }
  + L_{4,D}^2 { [ \eta\vert L_3\vert 3\ran \over [\eta 4] } \ .
\eeq
Therefore, the integrand in \eqref{boxini} becomes
\beq
{ N \over  L_{1,D}^2 L_{2,D}^2 L_{3,D}^2 L_{4,D}^2 }
\ , 
\eeq
with
\beq
N = \Big(  -\mu^2 { \lan 12\ran\over [12] }  + {X\over [12] } \Big)
     \Big( -\mu^2 { \lan 34\ran\over [34] }  + {Y\over [34] } \Big)\ .
\eeq
Thus we are led to consider the following four integrals:
\beqa
\label{remai}
  \cB_1  &= & \intD  { \mu^4\over L_{1,D}^2 L_{2,D}^2 L_{3,D}^2 L_{4,D}^2 }
             { \lan12\ran\lan34\ran\over [12][34] } \ ,
    \\ [8pt]
 \cB_2 &=&  - { \lan 12 \ran \over [12][34] } \intD { \mu^2 Y \over
                     L_{1,D}^2 L_{2,D}^2 L_{3,D}^2 L_{4,D}^2} \ ,
 \\[8pt]
\cB_3 &=& \cB_2\big( (1,2)\leftrightarrow(3,4) \big) \ ,
\\[8pt]
\cB_4 &=&  { \lan 12 \ran \lan 34\ran \over [12][34] } \intD { X Y \over
                    L_{1,D}^2 L_{2,D}^2 L_{3,D}^2 L_{4,D}^2 }
 \ .
\eeqa

The first one can be evaluated immediately:
\beq
\label{abo}
\cB_1\ =  \ {i\over (4 \pi)^{2-\e}} \, { \lan 12 \ran \lan 34\ran \over [12][34] } I_4^{4-2\epsilon}[\mu^4]
  \ =  \ {i\over (4 \pi)^{2-\e}}\,  K_4 \, { \lan 12 \ran \lan 34\ran \over [12][34] }
\ ,
\eeq
where $K$ is  defined in \eqref{funK}.

We will see that \eqref{abo} is actually the final result of the calculation, as all the additional contributions
will cancel out in the final expression.  In the $\e\to 0$ limit,  the  integral function $K$ is finite, see
\eqref{ffff}.

From this equation it is also clear that, in dimensional regularisation, the finiteness of
the all-minus amplitude  emerges from an $\e \times 1/ \e$ cancellation, where the $\epsilon$
factor is due to $\mu^2$ factors in the numerator of the integrand and the $1/\e$ is due to the
loop integration. 
As such, it is intriguing to speculate
\cite{Bardeen} that one may be able to interpret this amplitude as an anomaly.

The evaluation  of the remaining integrals in \eqref{remai}  is less immediate, and requires extensive 
use of PV reductions. Here
we will only quote the final result for the different expressions. We find%
\footnote{In the following expressions for the $\cB$'s we will omit an ubiquitous 
prefactor of $i/ (4 \pi)^{2-\e}$.} 
\beq
\label{coco}
\cB_2 + \cB_3 =
{ ([\eta | 12 | \eta] + [\eta | 34 | \eta])^2 \over [12] [34] \prod_{i=1}^{4} [\eta i]} {J_2 (t)\over t} \, + \,
 { [\eta | 12 | \eta]^2 + [\eta | 34 | \eta]^2 \over [12] [34]
  \prod_{i=1}^{4} [\eta i]} \bigg( -J_3 (s) + {2\over s} J_2 (s) \bigg)
\ ,
\eeq
and
\beq
\label{cococo}
\cB_4 =
 { [\eta | 23 | \eta]^2 + [\eta | 41 | \eta]^2 \over [23] [41]
 \prod_{i=1}^{4} [\eta i]} \,  \bigg( -J_3 (t) + {2\over t} J_2 (t) \bigg)
\, - \,
{ ([\eta | 23 | \eta] + [\eta | 41 | \eta])^2 \over [12] [34] \prod_{i=1}^{4} [\eta i]} {J_2 (t) \over t}
\ .
\eeq
Cancellation of the $J_2(t)$ terms in \eqref{coco} and 
\eqref{cococo} occurs thanks to the identity
$ [\eta | 23 + 41 - 12- 34| \eta]=0$, and we are left with
the following contribution from the box diagram:
\beqa
\label{bxfi}
\sum_{i=1}^{4} \cB_i &=& K_4 \, { \lan 12 \ran \lan 34\ran \over [12][34] }
+  { [\eta | 12 | \eta]^2 + [\eta | 34 | \eta]^2 \over [12] [34]
\prod_{i=1}^{4} [\eta i]}\bigg( -J_3 (s) + {2\over s} J_2 (s) \bigg)\, \\
 &+& \,
{ [\eta | 23 | \eta]^2 + [\eta | 41 | \eta]^2 \over [23] [41]
\prod_{i=1}^{4} [\eta i]} \,  \bigg( -J_3 (t) + {2\over t} J_2 (t) \bigg)
\nonumber
\ .
\eeqa

\subsection{The triangle diagrams}
In this section we consider the diagrams containing the one-loop three-point vertex we calculated earlier,
with the off-shell leg connected to a tree-level, three-point MHV vertex.
There are four such diagrams, one of which is depicted in figure 3. The other three diagrams are obtained by
cyclic permutation of the external particles. 
These  diagrams can be evaluated using our expression \eqref{final3}
for the three-point vertex, and the $- - +$ vertex. After a little algebra, one finds that the
triangle diagram in figure 3  gives
\beq
\cT_1\  = \
- {[\eta | 34  | \eta]^2 \over [12] [34]\prod_{i=1}^{4} [\eta i]}
\bigg( -J_3 (s) + {2\over s} J_2 (s) \bigg)\ ,
\eeq
and hence the sum over the four cyclic permutations gives
\beq
\label{trfi}
\sum_{i=1}^{4} \cT_1 \ = \
- { [\eta | 12 | \eta]^2 + [\eta | 34 | \eta]^2 \over [12] [34]
\prod_{i=1}^{4} [\eta i]} \bigg( -J_3 (s) + {2\over s} J_2 (s)
\bigg)\, -  \,
{ [\eta | 23 | \eta]^2 + [\eta | 41 | \eta]^2 \over [23] [41]
\prod_{i=1}^{4} [\eta i]} \,  \bigg( -J_3 (t) + {2\over t} J_2 (t) \bigg)
\ .
\eeq
Notice that when we sum the contribution of the triangle diagrams \eqref{trfi} to that of the boxes
\eqref{bxfi}, only the $K_4$ contribution survives, and we are left with
\beq
\sum_{i=1}^{4} \cB_i + \sum_{i=1}^{4} \cT_1  \ = \
 K_4 \, { \lan 12 \ran \lan 34\ran \over [12][34] }
 \ ,
 \eeq
which agrees with the expected result for the all-orders in $\e$ result for the four-point all-minus amplitude.
Next we have to consider diagrams which contain a one-loop bubble sub-diagram.
\begin{figure}
\begin{center}
$
\begin{array}{lcr}
\scalebox{0.3}{\includegraphics{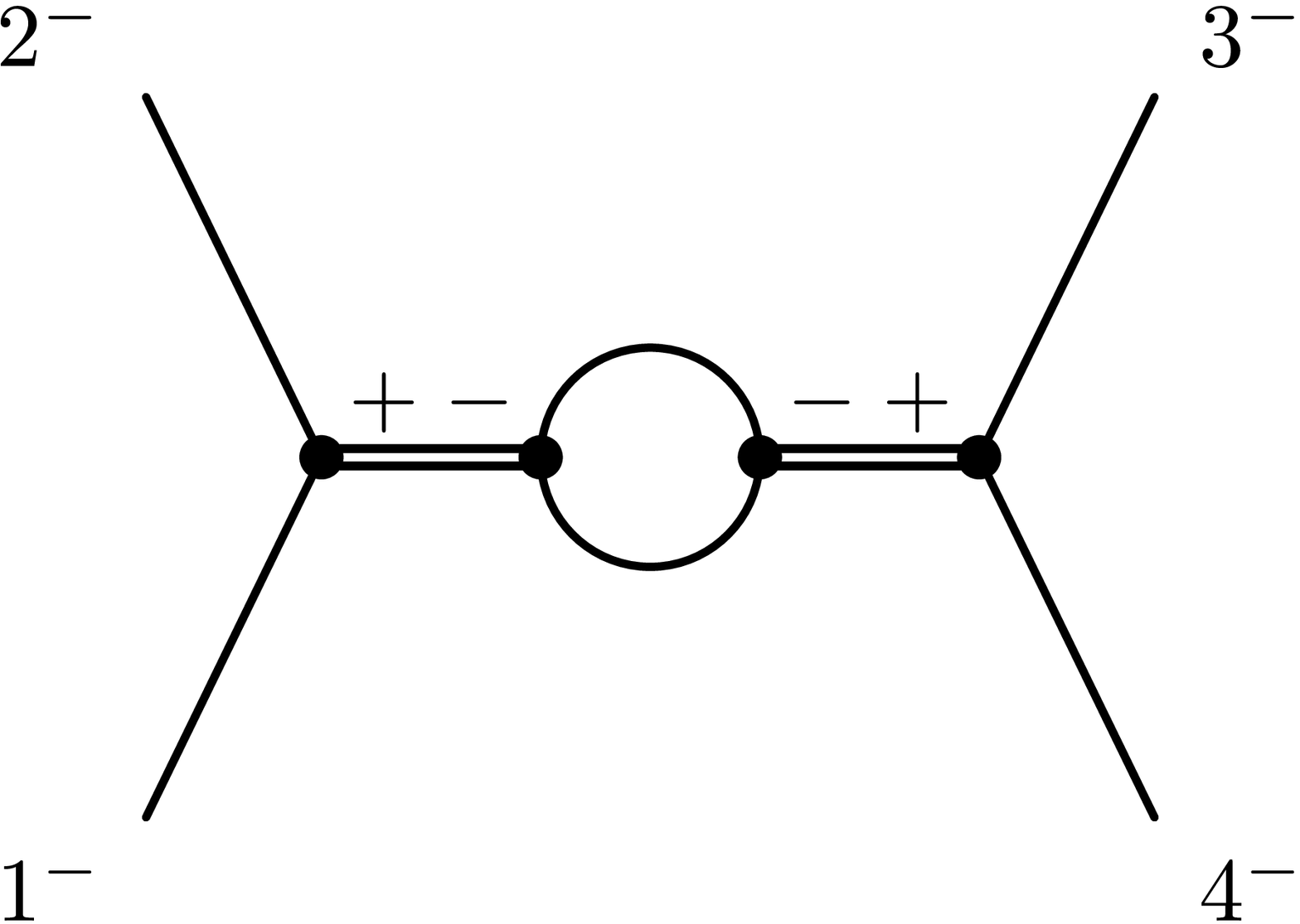}}
\label{trianglediagram}
& \;\;\;\;\;\;\;\; &
\scalebox{0.3}{\includegraphics{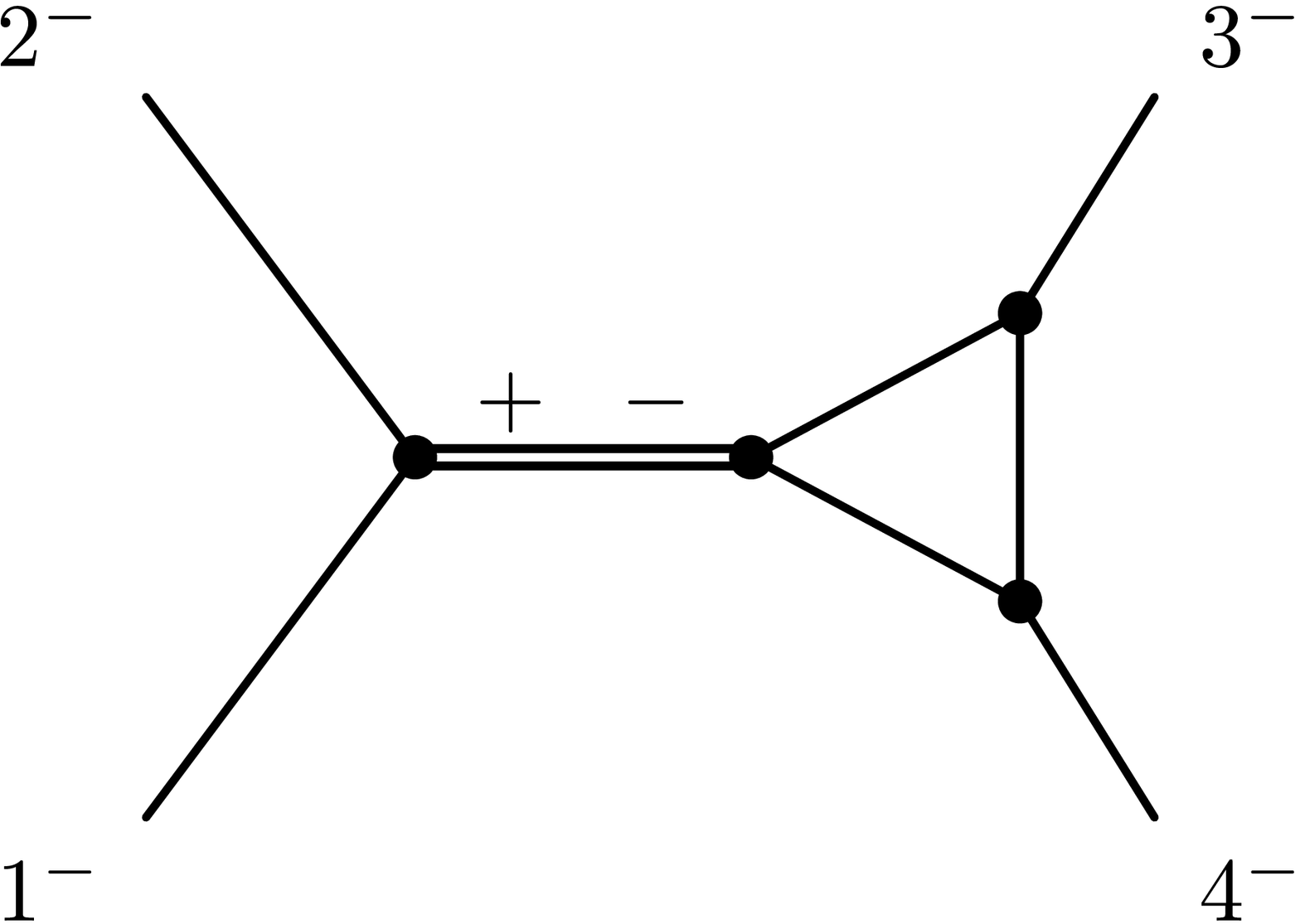}}
\end{array}
$
\caption{\it Bubble and triangle diagrams contributing to the all-minus four-gluon amplitude. The bubble diagram
is found to be zero. We also have to include three more triangle diagrams, obtained from the one in the figure
by cyclically rotating the external particles.}
\end{center}
\end{figure}

\subsection{The bubble diagram}
Finally we evaluate two bubble diagrams one of which is depicted in figure 3.
The bubble diagram from figure 3 gives
\beq
\label{zerobub}
\int\!{d^DL \over (2\pi)^D} \, {1\over L_{1, D}^2  L_{2, D}^2 }\, \Big({ [ \eta | L_1 |k \ran \over [\eta k]}\Big)^2
\ ,
\eeq
where $k$ is the lightcone truncation of
$K_{12} := k_1 + k_2$. The integral 
\eqref{zerobub}  is proportional to a two-tensor bubble integral
\beq
\label{notneeded}
\int\!{d^DL \over (2\pi)^D} \, {L_{1}^{\m} \, L_{1}^{\n}
\over L_{1, D}^2  L_{2, D}^2 }
\ ,
\eeq
whose expression is given in Appendix C.
However, the explicit expression is not needed. On general grounds
the tensor bubble is a linear combination of
$\eta^{\m \n}$ and $K_{12}^\m K_{12}^\n$.
Both tensor structures contract to zero when inserted into \eqref{zerobub}. Since the
other bubble diagram obtained by cyclic permutation of the external lines is also zero,
this class of diagrams gives a vanishing contribution.


\section{A holomorphic field redefinition}
In this section we  consider a particular holomorphic change of variables in the
Yang-Mills path integral,%
\footnote{Also recently studied in \cite{stony}.} 
which is different from the canonical field redefinition of
Mansfield \cite{Mansfield}.
The special form of the Chalmers-Siegel Lagrangian \eqref{cs},
which we write here for convenience as 
\beq
\label{pb}
\cL_{\rm CS} := {1 \over 2} \bar{A} \Big(  \Box \, A +2  i [
\del^- A , \del^z A ]  \Big) \ ,
\eeq
suggests the choice of a new set of variables $(B,\bar{B})$, where $\bar{B} = \bar{A}$ and
$B$ is a function of $A$ alone such that the Lagrangian is free, when written in terms of
$(B,\bar{B})$,
\beq
\label{pata}
\cL_{\rm CS} := {1 \over 2} \bar{B} \Box \, B
\ .
\eeq
In this sense, this change of variables is similar to 
 the B\"{a}cklund transformations which have been applied to Liouville  theory
 \cite{Thorn}. 
 
Equations  \eqref{pb} and \eqref{pata}
determine $A=A(B)$ to be the solution of
\beq
\label{fmo}
 \Box \, A +2  i [ \del^- A , \del^z A ] \ = \ \Box B
 \ .
 \eeq
 This change of variables is non-canonical, and only transforms the
 $A$ fields, leaving the $\bar{A}$ field untouched. Moreover
 it is holomorphic, $A= A(B)$ so that $B$ has
 no dependence on the $\bar{A}$ fields.
 One can solve \eqref{fmo} perturbatively,
 \beq
 A \ = \ A^{(0)} \, + \, A^{(1)} \, + \, A^{(2)} \, + \, \cdots
 \ ,
 \eeq
 where
 $A^{(0)} = B$, $A^{(1)} = -2i \Box^{-1}  [ \del^- B , \del^z B ]$, $\ldots$
In general $A^{(n)}$ contains $n+1$ insertions of the field $B$.
Interestingly, the solution of the self-dual Yang-Mills equations of motion can be
bootstrapped using a Bethe Ansatz \cite{Bardeen,Cangemi,Korepin}
and has intriguing  connections to integrability.

If this change of variables is applied to the full Yang-Mills Lagrangian,
a full set of MHV-like vertices is generated. One can readily see that
these cannot be equal to the known MHV vertices, as derived explicitly
in \cite{Ettle:2006bw} using a non-holomorphic but canonical
change of variables. The reason is that certain contributions, that are needed
to get the full MHV vertices as in \cite{Ettle:2006bw}, are
missing in the expressions obtained from the holomorphic field redefinition.
However, one expects the same on-shell expressions, and very recent
investigations appear to confirm this \cite{stony}.

Specifically, the fact that we transform the field $A$ but not $\bar{A}$ has an important
consequence. Consider the four-point gluon vertex. In terms of the new fields,
it receives two kinds of contributions: {\bf a.} a contribution from the
original four-point vertex in the Lagrangian, where we substitute
$\bar{A} = \bar{B}$ and the lowest-order term in
the expansion of the solution $A=A(B)$ to \eqref{fmo}, $A^{(0)} = B$; and
{\bf b.} a contribution from the three-point vertex in the Yang-Mills
Lagrangian,  of the form (schematically) $\bar{A}\bar{A} A$, where we
replace $A$ by the term in the expansion of $A=A(B)$ containing two $B$-fields.
The term we generate is thus of the form%
\footnote{A trace over colour  is always understood.}
$BB \bar{B} \bar{B}$. Importantly, this correction alters
the vertex with helicities $--++$, but not the split-helicity
vertex  $-+-+$. This means that the perturbative expansion
in terms of the new fields is going to be different from that giving rise to MHV rules.

\subsection{The all-plus helicity amplitude from a Jacobian}
 We now discuss how the new field variables defined above yield
 the all-plus helicity amplitudes.
 The holomorphic change of variables is not canonical and 
 leads to a nonvanishing Jacobian
 \beq
 \label{jj}
 \cJ (B) \ = \  {\rm det_{\it x,y}} \left( {\d A (x) \over \d B(y)}\right) (B)
\ ,
\eeq
which is a functional of the $B$ fields only, and as such can contribute to the
all-plus scattering amplitude, obtained from the
correlator $\lan \bar{A} (x_1) \cdots \bar{A} (x_n) \ran$
upon application of standard reduction formulae.
More precisely, we wish to compute
\beq
\label{allabar}
G(x_1, \ldots , x_n) \ := \
\lan \bar{A} (x_1) \cdots \bar{A} (x_n) \ran \ = \
\lan \bar{B} (x_1) \cdots \bar{B} (x_n)\, \cJ^{(n)} (B)  \ran_{\rm free}
\ ,
\eeq
where in the right hand side we perform free Wick contractions of the $B$ and $\bar{B}$ fields, and
$\cJ^{(n)} (B) $ is the term in the Jacobian \eqref{jj} which contains $n$ $B$ fields.

Now we argue that in this way we precisely obtain the all-plus amplitude,
which is missing within the MHV diagram formalism. Our strategy is very simple.
We calculate the Jacobian and its contribution to the all-plus correlation function in terms
of diagrams (without evaluating them explicitly),
and show that these diagrams precisely match those obtained with lightcone quantisation
(these are the parity conjugate diagrams of those considered in the all-minus calculation).

This is most easily seen in a toy model, which captures the essence of the
problem. In this model, the cubic interaction of self-dual Yang-Mills is replaced by
the simpler cubic interaction $\l \bar{A} A A$ which is free of derivatives. We  consider
\beq
\cL := \bar{A} ( - \Box A  + \l A^2)
 \ ,
 \eeq 
 and we redefine fields so that
 \beq
 \cL = - \bar{B} \Box B
\ .
\eeq
We use the holomorphic change of variables
\beq
\label{toyyeq}
\bar{A}\ = \   \bar{B}
\ ,
\eeq
 and
\beq
\label{toyeq}
 - \Box A  + \l A^2\ =\ - \Box B
  \ .
 \eeq
The three-point vertex of the original Lagrangian
is proportional to  $-\l f^{abc}$ and contributes to the Green function
$\lan \bar{A}(x) \bar{A}(y) A(z) \ran$.

At the level of functional integrals, we have the equalities
\beq
\label{Z}
Z(J, \bar{J}) :=
\int \! \delta A \delta \bar{A}\  e^{ - S( A, \bar{A}) + \int \bar{J}A + \int \bar{A} J}
=
\int \! \delta B \delta \bar{B} \ \cJ (B) \, \  e^{ - S_{\rm free}( B, \bar{B}) +
\int \bar{J}A(B)  + \int \bar{B} J}
\ .
\eeq
Notice the presence of the Jacobian $\cJ(B)$ defined in \eqref{jj}, and the fact that the source
$\bar{J}$ is coupled to $A = A(B)$,  i.e.~the solution to the change of variables
\eqref{toyeq}.
Taking a functional derivative of \eqref{toyeq}  with respect to $B$ we get
\beq
\Big(-\Box_x + 2 \l A(x) \Big) {\d A(x) \over \d B(y)} = - \Box \d (x-y)
\ ,
\eeq
or, formally,
\beq
{\d A(x) \over \d B(y)}  \ = \ (1 - 2\l  \Box^{-1} A)^{-1}_{x,y}
\ .
\eeq
We have
\beqa
\label{cucu}
\cJ & := &  {\rm det}_{x,y} \left({\d A(x) \over \d B(y)} \right)  = e^{{\rm Tr} \log (1 - 2\l  \Box^{-1} A)^{-1}}
\ = \ e^{- {\rm Tr} \log (1 - 2\l  \Box^{-1} A) }
\\ \nonumber
&=& \exp \left[ {\sum_{m=1}^{\infty}  {1 \over m} {\rm Tr} \big[ (2\l \, \Box^{-1} A)^m \big]}\right]
\ ,
\eeqa
where%
\footnote{\eqref{tr}  is meant to be properly regularised, by e.g.~using dimensional regularisation.}
\beq
\label{tr}
{\rm Tr} \big[ (2 \l \, \Box^{-1} A)^m \ = \ (2 \l)^m \int\! d^4y_1 \cdots d^4y_m \ G(y_1 - y_2) A(y_2) \,
G(y_2 - y_3) A(y_3) \cdots G(y_m - y_1) A(y_1)
\ , 
\eeq
and $\Box G(x) = \d^{(4)}(x)$. 
Our correlator \eqref{allabar} becomes then
\beqa
\label{fifi}
&& G(x_1 , \ldots , x_n) \ = \
\\ \cr
&&
\left \lan \bar{B} (x_1) \cdots \bar{B} (x_n)\,
 e^{  { \sum_{m=1}^{\infty}  {(2\l )^m \over m}
 \int\! d^4y_1 \cdots d^4y_m \ G(y_1 - y_2) A(y_2) \,
G(y_2 - y_3) A(y_3) \cdots G(y_m - y_1) A(y_1)}
}
\right\ran_{\rm free}
\ ,
\nonumber
\eeqa
where on the right hand side of \eqref{fifi} (as well as \eqref{cucu})
one should think of $A$ as the functional of the $B$-fields
$A= A(B)$ given by the solution of
\eqref{toyeq}. Moreover, we need to pick the term in the expansion of
the exponential which contains $n$ $B$-fields, as  indicated
on the right hand side of \eqref{allabar}.

In order to determine $A=A(B)$, we need to solve \eqref{toyeq}, and we can do that perturbatively.
We set
  \beq
  \label{simon}
  A(B) = A^{(0)} +A^{(1)} +A^{(2)} + \cdots
   \ ,
   \eeq
   where we find $A^{(0)} = B$, and
   \beq
   \label{fite}
   A^{(1)} = \l \, \Box^{-1} B^2
    \  ,
    \eeq
and so on.

Now we need to perform free Wick contractions in \eqref{fifi}.
If we pick the solution to lowest order, $A^{(0)} = B$,
we obtain  a contribution  proportional to
\beq
\l^n
 \int\! d^4y_1 \cdots d^4y_n\ G(y_1 - y_2) \,
G(y_2 - y_3)  \cdots G(y_n - y_1) \ G(x_1 - y_1) \cdots G(x_n - y_n)
\ .
\eeq
This is nothing but the $n$-vertex polygon diagram one would draw with the
$\l \bar{A} AA$ vertex in the original Lagrangian; at four points, it is
(the parity conjugate of) the box diagram in the all-minus calculation --
the diagram with the maximum number of propagators, $n$.

The diagrams corresponding, in our four-point example, to triangle and bubble
topologies, arise by taking precisely one insertion of $A^{(1)}$ and two of $A^{(0)}$,
and two insertions of $A^{(1)}$ for the fields $A$ appearing on the right hand side
of \eqref{fifi}, respectively.
Let us check this in more detail for the triangle diagram in the four-point case.
The corresponding contribution is proportional to
\beqa
\label{untr}
&&
\l^4\int\! d^4y_1d^4y_2 d^4y_3\ G(y_1 - y_2) \,
G(y_2 - y_3)  G(y_3 - y_1) \ G(x_1 - y_1) G(x_2 - y_2) \,
\nonumber \\
&&\qquad \qquad\qquad \qquad
\Box^{-1}_{y_3} \left[ G( x_3 - y_3) G(x_4-y_3) \right]
\ ,
\eeqa
and  matches the triangle diagram in figure 3
(we recall that $\Box^{-1}_{y } f (y) = \int\! d^4z \, G(y-z) f(z)$).
Indeed, the insertion of $\Box^{-1}$ in \eqref{untr} corresponds to the propagator connecting the
tree-level vertex in the figure to the loop, and the two propagators
$G( x_3 - y_3) G(x_4-y_3)$ are attached to this tree-level three-point vertex.
The propagators $G(x_1 - y_1) G(x_2 - y_2)$ are those emerging from the loop.

The bubble-like diagram would contribute
\beq
\label{unbub}
\l^4\int\! d^4y_1d^4y_2 \ \Box^{-1}_{y_1} \left[ G( x_1 - y_1) G(x_2-y_1) \right] \,
\Box^{-1}_{y_2} \left[ G( x_3 - y_2) G(x_4-y_2)\right]
\ ,
\eeq
again matching the parity conjugated of the bubble-diagram (see figure 3) in the four-minus calculation
for the toy model.

The analogous calculation for the self-dual Yang-Mills theory would clearly
reproduce  the same  diagrammatics, and would be expected to
give the correct expressions for the amplitudes.

\subsection{The single-minus and the single-plus helicity amplitudes}

The amplitudes in pure Yang-Mills where all gluons but one have the same helicity are quite special
because of their finiteness, a feature in common with the all-plus and the all-minus amplitudes.
In this section we would like to sketch how they are calculated within the framework of the
holomorphic change of variables introduced above.

The amplitude with a single positive helicity and its parity conjugate, the amplitude with a single
negative helicity, have a quite different origin, so we will discuss them separately.
In each case, the strategy will be to map the contributions arising from the change of variables
to those of lightcone Yang-Mills perturbation theory.

\subsubsection{The single-plus amplitude}
The $n$-point single-plus amplitude must come from a diagram containing
precisely $n-1$ MHV vertices.%
\footnote{An amplitude with  $q$ negative-helicity gluons is built out of $q-1+L$ MHV vertices,
where $L$ is the number of loops.}
The correlation function corresponding to such amplitude is
\beq
G(x_1, \ldots, x_n) \ := \ \left\lan
A(x_1) \cdots A(x_{n-1}) \bar{A}(x_n) \right\rangle
\ ,
\eeq
which we would like to evaluate using the new variables.
Here we will limit our attention to 
the $n$-gon diagram  contribution, in the simple case
of $n=4$. Generalisation to $n>4$ particles and 
diagrams with other topologies is straightforward.

The relevant term to study, after re-expressing $G(x_1 , \ldots , x_n) $ in terms of
the new fields,  is
\beq
\label{nn}
\Big\lan
A(x_1)  A(x_2) A(x_{3}) \bar{B}(x_4) \,
\l^3 \int\! \prod_{i=1}^3 d^4z_i \ ( \bar{B}\bar{B} A)(z_1)\,
( \bar{B}\bar{B} A)(z_2)\,( \bar{B}\bar{B} A)(z_3)\,
\ \cJ (B)
\Big\ran_{\rm free}
\, ,
\eeq
where $A=A(B)$ and the subscript ``free" instructs one to take free Wick contractions
of the $B$ and $\bar{B}$ fields.
Notice that in \eqref{nn} the old fields $A=A(B)$, expressed as functionals of the new fields,
are inserted.
A na\"{i}ve application of the equivalence theorem would allow us to replace
the old fields by the new ones in a correlation function, without changing the
corresponding scattering amplitudes (the correlation function would
of course be different). In the case of the holomorphic field redefinition discussed above,
the equivalence theorem cannot be applied, because that field redefinition
is singular precisely on the mass shell. This issue is discussed
in Appendix \ref{equivalence}.%
\footnote{The equivalence theorem has a long history. For a back of the envelope
proof, see \cite{coleman}.
The issue of a violation of the equivalence theorem using the change of
variables discussed here was also recently addressed in
\cite{stony}.}

In order to generate a box-diagram structure, we need to open up an extra propagator.
This is obtained from taking in any of the three insertions of the
functional $A(B)$ (determined by \eqref{toyeq}) the first nontrivial iteration in the solution
$A^{(1)}$, calculated in \eqref{fite}, and for the remaining $A$
fields, the zeroth order iteration, $A^{(0)} = B$.
A typical term will look like
\beqa
\nonumber
&&
\l^4 \Big\lan
B(x_1)  B(x_2) B(x_{3}) \bar{B}(x_4) \,
 \int\! \prod_{i=1}^3 d^4z_i \ ( \bar{B}\bar{B} B)(z_1)\,
( \bar{B}\bar{B} B)(z_2)\,
\\
&&
\hspace{3.5cm}
\times\ ( \bar{B}\bar{B} ) (z_3)\, \int\! dz^4_4 \, G(z_3 - z_4) (B B) (z_4)
\Big\ran_{\rm free}
\, .
\eeqa
Notice that we have picked the zeroth order term in the Jacobian, as well as for the external
field $A(B)(x_4)$.
Contracting  $B(z_4)$ with $\bar{B}(z_1) $, $B(z_1)$ with $\bar{B}(z_2) $, and
$B(z_2)$ with $\bar{B}(z_3) $, we obtain the required box structure, proportional to
\beq
\l^4 \int\!\prod_{i=1}^{4} d^4z_i \ G(z_1-z_2) G(z_2-z_3) G(z_3-z_4) G(z_4-z_1)
\ .
\eeq
The remaining fields which are integrated over contract in an obvious way with the insertions
$B(x_1)$,  $B(x_2)$,  $B(x_{3})$,  and $\bar{B}(x_4) $.

\subsubsection{The single-minus amplitude}
Finally, let us consider the single-minus helicity amplitude.
The relevant correlation function is
\beq
G(x_1, \ldots, x_n) \ := \ \left\lan
\bar{A}(x_1) \cdots \bar{A}(x_{n-1}) A(x_n) \right\rangle
\ .
\eeq
Again, we focus only on the four-point case, and specifically on the box -function contribution.
We expect to use a single MHV three-point vertex; the relevant box-like term is obtained from
\beq
\label{cosi}
\left\lan
\bar{B}(x_1) \bar{B}(x_2) \bar{B}(x_3)A(x_4)  \ \l \int\!d^4z (\bar{B} \bar{B} A)(z) \, \cJ(B)
\right\rangle_{\rm free}
\ .
\eeq
The box diagram structure clearly comes from considering the $\l^3$ iteration of
the field $A(B)$ which sits inside the $z$-integral. This iteration is schematically of the form
\beqa
\label{dd}
A^{(3)}& \sim&    \l \Box^{-1} (A^{(0)}A^{(2)} + A^{(2)}A^{(0)}+ A^{(1)}A^{(1)}  )
\\ \nonumber
&\sim&
\l^3 (\Box^{-1} B \Box^{-1} B \Box^{-1} B^2 ) \ + \ \l^3 (\Box^{-1} B^2) (\Box^{-1} B^2)
\ .
\eeqa
In order to make a box diagram we need three integrations, so we need to pick the first term
in \eqref{dd}. We rename $z\to z_1$ and use
 \beq
(\Box^{-1} B \Box^{-1} B \Box^{-1} B^2 )  (z_1)
 =
\int\! \prod_{i=2}^{4} d^4z_i \ G(z_1-z_2)B(z_2) \, G(z_2-z_3) B(z_3) \, G(z_3-z_4)(BB)(z_4)
\ .
\eeq
The fourth propagator needed to form a box is obtained from contracting
$B(z_4)$ with one of the two $\bar{B}(z_1)$ in \eqref{cosi}.
Finally, the remaining fields are contracted in an
obvious way with the insertions $\bar{B}(x_1)$, $ \bar{B}(x_2)$,  $\bar{B}(x_3)$,
and $B(x_4)$. The box diagram arises from setting
$A(x_4) \to B(x_4)$, and from taking the trivial contribution $\cJ \to 1$ to the Jacobian.

\subsection{Perturbation theory in the new variables}

Now consider the question of perturbation theory in the new variables $(B,\bar{B})$.
As we have seen there are three types of contributions -- those coming directly from the
Lagrangian, expressed in terms of the new variables, those coming from the Jacobian, and finally
those coming from violations of the equivalence theorem, which express themselves through
contributions from field redefinitions of the external states. We have argued that the
Jacobian terms alone give rise to the one-loop all-plus amplitudes, while the all-minus 
is obtained solely from three-point MHV vertices, which are not modified in our 
holomorphic field redefinition. 
Our analysis indicates that the single-minus and the single-plus amplitudes arise from combinations 
of MHV-type vertices with the Jacobian  and external state contributions.
In general, the procedure for calculating complete amplitudes with generic 
helicities using these new variables is then clear: one
combines the effective vertices from the Lagrangian with the equivalence theorem violating
contributions and those arising from the  Jacobian.


\section{Conclusions}

We would like to end with some comments and speculations:

{\bf 1.}
Given the fact that the MHV three-point vertices are nothing but the lightcone vertices,
the agreement of our MHV diagram calculation with the  all orders in $\e$ result
for this amplitude will 
persist for all $n$-point all-minus  scattering amplitudes.

{\bf 2.} The fact that the all-minus amplitude is non-zero arises from
an $\e \times1/ \e$ cancellation, as we showed explicitly in the calculations of sections
3 and 4.
This is reminiscent of an anomaly, as also noted by numerous authors in
the context of related calculations \cite{Bardeen,Cangemi,Mahlon1,Mahlon2,csanom}.
It would be interesting to make this statement more precise, 
and to understand which symmetry is anomalous.

{\bf 3.} We have explained the all-plus amplitude as coming from a Jacobian.
This applies to our holomorphic change of variables,
which is different from Mansfield's canonical transformation \cite{Mansfield}.
However, it is likely that a proper quantum treatment of Mansfield's approach
would also lead to a Jacobian which then would yield the all-plus amplitude.

{\bf 4.} After arguing that we can derive the all-plus amplitudes, we discussed how
to obtain the single negative/positive helicity amplitudes in a toy model which is
closely related to the Yang-Mills case. Clearly it would be of interest to do this directly
in Yang-Mills.

{\bf 5.} In supersymmetric theories the amplitudes with zero or one particle of one helicity
type vanish, due to supersymmetric Ward identities. Thus the Jacobians  must 
cancel  in these cases. It would be interesting to show this explicitly.

{\bf 6.} Finally,  it is clear  that the holomorphic change of 
variables discussed in section 5 gives a different perturbation theory. 
But one can ask whether this yields practical rules, 
or, more precisely, if  the perturbative expansion can be reassembled into an effective set of vertices 
which incorporate equivalence theorem violating contributions
and possibly contributions  from the Jacobian.
This seems to us an interesting avenue to follow.


                \section*{Acknowledgements}

It is a pleasure to thank Zvi Bern, Rutger Boels, Freddy Cachazo, Paul Heslop, 
Paul Mansfield, Tim Morris, Sanjaye Ramgoolam,  Diana Vaman and Costas Zoubos  
for discussions, and
Simon McNamara for collaboration at the initial stage of this project.
We would like to thank PPARC for support under a
Rolling Grant PP/D507323/1 and the Special Programme Grant PP/C50426X/1.
The work of GT is supported by an EPSRC Advanced Fellowship and
by an EPSRC Standard Research Grant.

\newpage

\startappendix

\Appendix{A brief review of lightcone gauge theory}
\label{lcym}

We begin by introducing lightcone coordinates
\beq
 x^{\pm} :=  {x^0 \pm x^3  \over
\sqrt{2}} \ , \qquad x^{\hat{\m}} :=  (x^2, \ldots , x^D) \ .
\eeq
In terms of these, the scalar products between two vectors $A$ and $B$ is
\beq
A \cdot B := A^+ B^- +
A^- B^+ - A^{\hat\m} B^{\hat\m}
\ , 
\eeq
where $\hat\m$, $\hat\n=1,2$ and we defined
$A^{\hat\m} B^{\hat\m} :=  A^{\hat\m} B^{\hat\n}\d_{\hat\m \hat\n}$.

The lightcone gauge is defined by
\beq
\label{lc} A^- = 0 \ .
\eeq
Equivalently, we can write the above condition as $\eta \cdot A =0
$, where $\eta$ is a constant null vector, chosen to have components  $\eta := ( 1,0, 0,1)$.

In the lightcone gauge, the Yang-Mills Lagrangian is%
\footnote{In this and the following Lagrangians we will omit an obvious trace over the group
generators to simplify the notation. }
\beqa
\label{lym}
\nonumber
\cL_{\rm{YM}} &:= &-{1\over 4} F_{\m
\n}F^{\m \n} = {1\over2} (\del^- A^{+})^2 - (\del^- A^{\hat\m})(\del^{\hat\m} A^+) - i A^+ [
\del^- A^{\hat\m} , A^{\hat\m}] + (\del^- A^{\hat\m})(\del^+ A^{\hat\m})
\\
&-&{1\over 4} F_{\hat\m
\hat\n}F^{\hat\m \hat\n} \ .
\eeqa
From this equation it is clear that $A^+$ is a Lagrange
multiplier, and can be integrated out. This amounts to replacing it by the solution to the
equations of motion, which is
\beq
A^+ = (\del^-)^2 ( \del^{\hat\m} \del^- A^{\hat{\m}} + i
[A^{\hat\m},  \del^- A^{\hat\m}   ] ) \ .
\eeq
Plugging the solution back into \eqref{lym}, one
arrives at
\beq \label{lym2}
\cL_{\rm YM} = (\del^- A^{\hat\m} )(\del^+ A^{\hat\m})  - {1\over 4}
F_{\hat\m \hat\n}F^{\hat\m \hat\n} + {1\over 2} ( \del^{\hat\m} \del^{-} A^{\hat \m} - i
[\del^- A^{\hat\m} , A^{\hat\m}] ) (\del^-)^2 ( \del^{\hat\n} \del^{-} A^{\hat \n} - i [\del^-
A^{\hat\n} , A^{\hat\n}] ) \ .
\eeq
We now specialise to the four-dimensional case, introducing
the two complex combinations
\beq \label{azs}
 A^z := {A^1 + i A^2 \over \sqrt{2} } \ , \qquad
A^{\bar{z}} := {A^1 - i A^2 \over  \sqrt{2} } \ ,
\eeq
which are the fields for gluons of
positive and negative helicity, respectively. The change of variables \eqref{azs} leads to a
remarkable simplification in the structure of the four-dimensional Yang-Mills Lagrangian,
converting \eqref{lym2} into
\beq
\label{lymlc} \cL:= \cL_{-+} +\cL_{-++}+\cL_{+--} +
\cL_{--++} \ ,
\eeq
where $\cL_{-+} = - A_{\bz} \Box A_z$ is the free term, with $\Box := 2 (
\del_+ \del_-  - \del_z \del_{\bz})$, and
\beqa
\cL_{-++} & = &  2 i [ A_z , \del_{+}  A_{\bz}
] (\del_+)^{-1} ( \del_{\bz} A_z )
 \ ,
 \\ \nonumber
 \cL_{+--} & = &  2 i [ A_{\bz} , \del_{+}  A_{z} ] (\del_+)^{-1} ( \del_{z} A_{\bz} )
 \ , 
\\ \nonumber
\cL_{--++} & =&  -2  [A_{\bz} , \del_{+} A_z ] (\del_+)^{-2}  [ A_z , \del_{+}  A_{\bz} ] \ .
\eeqa
 This form of the Yang-Mills Lagrangian precisely agrees with that in Eq.~7 of
\cite{Chalmers:1998jb}.%
\footnote{After replacing $\Box \to - \Box$ because of different conventions for the metric,
and  multiplying their Lagrangian by an overall factor of -2.}
 We also remark that the
combination
\beq
\cL_{\rm SDYM} := \cL_{-+} +\cL_{-++}
\eeq
describes  the self-dual truncation
of pure Yang-Mills theory (the combination $ \cL_{-+} +\cL_{--+}$ describes the anti-self-dual
truncation).

Upon making the change of variables $\phi := (\del^-)^{-1} A_z$, $G^{z-} := -2
\del^- A_{\bz}$, it becomes
\beq
\label{cs}
\cL_{\rm CS} := {1 \over 2} G^{z-} \Big(  \Box \, \phi +2  i [
\del^- \phi , \del^z \phi ]  \Big) \ ,
\eeq
 which is the Chalmers-Siegel Lagrangian for
self-dual Yang-Mills  \cite{Chalmers:1996rq}. This is easily derived in a first-order
formulation, i.e.~by starting from $\cL := (1/2) G_{\m \n} F^{\m \n}$, where $G_{\m \n}$ is an
anti-self-dual field strength, and further imposing the lightcone gauge condition \eqref{lc}.
\Appendix{On  the equivalence theorem}
\label{equivalence}

The purpose of this appendix is to show that when
we ``redefine away" the three-point vertex
in the Chalmers-Siegel Lagrangian \eqref{pb} 
via the holomorphic change of variables 
defined by
$\bar{B} =    \bar{A}$, and
$ \Box A +2  i [ \del^- A , \del^z A ]  =  \Box B$,
the three-point interaction reappears as a violation of
the equivalence theorem.  This theorem can be briefly stated
as follows. Consider a theory defined by an action $S(\phi)$ and functional integral
\beq
\label{primo}
Z(J) \ := \ \int\!\d\phi \ e^{-S(\phi) + \int\! dx \, J(x) \phi(x)}
\ .
\eeq
Now consider a different functional integral
\beq
\label{secondo}
\tilde{Z}(J) \ := \ \int\!\d\phi \ e^{-S(\phi) + \int\! dx \, J(x) \varphi ( \phi) (x)}
\ ,
\eeq
where $\varphi :=\varphi( \phi )$ are  new fields defined by an invertible change of variables,
with $\varphi (\phi)  = c \phi + \cO (\phi^2)$, and $c$ is a constant different from zero.
The Green functions obtained from \eqref{primo} and \eqref{secondo} are clearly different.
However, if $\varphi= \varphi (\phi)$ are good interpolating fields
for the quanta carried by $\phi$,
scattering amplitudes of the theories \eqref{primo} and \eqref{secondo}
are  identical modulo a wave-function renormalisation, which compensates
a possible non-canonical normalisation of the new fields.

Notice finally that one  could recast \eqref{secondo} as
an integral over the new fields,
\beq
\label{terzo}
\tilde{Z}(J) \ := \ \int\!\d\varphi \  \left| \left| {\d \phi \over \d \varphi} \right| \right|
\,
e^{-\tilde{S}(\varphi) + \int\! dx \, J(x) \varphi (x)}
\ ,
\eeq
where $\tilde{S}(\varphi) := S(\phi (\varphi))$ is the action
in terms of the new fields.

We now consider the issue of a possible violation of the equivalence theorem
using our toy model, defined by the action
\beq
\label{sa}
\cL := \bar{A} ( - \Box A  + \l A^2)
 \ .
 \eeq
We then redefine fields holomorphically using \eqref{toyyeq} and
\eqref{toyeq},
so that
 \beq
 \label{vo}
 \cL = - \bar{B} \Box B
\ .
\eeq
Now we would like to calculate at tree level the correlator,
\beq
\label{tre}
\lan \bar{A}(x) \bar{A}(y) A(z) \ran=
{\delta^3 Z\over \delta J(x) \d J(y) \d \bar{J}(z)}
\ ,
\eeq
which we evaluate using the new-fields representation of
the functional integral $Z$ defined in \eqref{Z}.
If we could apply the equivalence theorem, we would be able to replace
$A(B)$ with $B$ on the right hand side of \eqref{Z}. This would lead to the 
wrong conclusion that the  Green function \eqref{tre} vanishes
because the action $S_{\rm free}(B, \bar{B})$ is free.
Instead we use the expansion of the old field $A$ as 
a function of $B$ as in \eqref{simon}, and  find 
\beq
\label{ciccio}
\lan \bar{A}(x) \bar{A}(y) A(z) \ran \ = \
\lan \bar{B}(x) \bar{B}(y) \l \Box^{-1} B^2 (z) \ran_{\rm free}\, +\,
\lan \bar{B}(x) \bar{B}(y) B(z) \, 2 \l {\rm Tr} ( \Box^{-1} B  )  \ran_{\rm free}
\ .
\eeq
The first term on the right hand side of \eqref{ciccio}
comes from the fact that the sources
are coupled to the old fields. The
equivalence theorem would imply that its contribution to the correlator
is nonzero, but vanishes upon LSZ reduction i.e.~does not contribute to
the scattering amplitude.
The second term comes from
a nontrivial Jacobian. The Jacobian term contains a trace
which must be properly regularised (it entails an integral over
positions). However the trace is also over colour, hence it gives
either zero or a subleading term in the large-$N$ limit. We will discard it.%
\footnote{Moreover, it is a loop effect, and we are now focussing our attention 
only on the tree-level contribution to \eqref{tre}.}
The first term is nonvanishing and we get (we  omit an $f^{abc}$ in front):
\beqa
\label{ciccio2}
\lan \bar{A}(x) \bar{A}(y) A(z) \ran & = &
\l \, \Box^{-1}_{z} \Big( G(z-x) G(z-y) \Big)
\\ \nonumber
&=&
\l \int\! d^4t \ G(x-t) G(y-t) G(z-t)
\ ,
\eeqa
where $G(x)$ is a free scalar propagator.
This is the term which, according to the equivalence theorem, should
contribute zero to the scattering amplitude. This is clearly not the case here.
To obtain the contribution to the scattering amplitude, we
apply LSZ reduction. The first step consists in  amputating propagators
on the external legs. This is achieved by multiplying
$\lan \bar{A}(x) \bar{A}(y) A(z) \ran $
by
$\Box_x \Box_y \Box_z$.
We get
\beq
\Box_x \Box_y \Box_z \
\lan \bar{A}(x) \bar{A}(y) A(z) \ran =
\l \, \d(x-z)\d (y-z)
\ .
\eeq
In the second step we  multiply by the wave-functions of free particles,
$\exp (i p_1 x + i p_2 y + i p_3 z)$, setting $p_1^2 = p_2^2 = p_3^2 = 0$,
and Fourier transform  in $x,y,z$.
The resulting three-point scattering amplitude is
\beq
\label{three}
\l \,\int\! d^4x\, d^4y \, d^4z \
\d(x-z)\d (y-z) e^{i p_1 x + i p_2 y + i p_3 z}
\ = \ \l \ (2 \pi)^4 \d^{(4)} (p_1 + p_2 + p_3)
\ .
\eeq
This is nonvanishing when $p_1^2 = p_2^2 = p_3^2 = 0$,
and is precisely the contribution of the
tree-level three-point vertex in the original Lagrangian for the
$A$ and $\bar{A}$ fields.

Our final remark is that this violation of the equivalence theorem is entirely expected.
Indeed, starting from the canonically normalised field $A$, the transformation
\eqref{toyeq} is singular, in the sense that it involves the operator
$\Box^{-1}$, which is singular precisely on the mass shell, where scattering amplitudes
are computed.  Hence, the field redefinition $A= A(B)$ in  \eqref{toyeq}
cannot be compensated simply by a wave-function renormalisation
(equivalently, the fields $B$, $\bar{B}$ are
not good interpolating fields for the quanta carried by $A$, $\bar{A}$).

\Appendix{Integrals}

For convenience, we here summarise some integrals used in this paper.

The scalar $n$-point integral functions
in $D=4+2m-2 \eps$ dimensions are defined as
\beqa
I^{D}_{n} \equiv I^{D}_{n}[1] & = &
i (-1)^{n+1}(4 \pi)^{D/2} \int \frac{d^{D}L}{(2 \pi)^{D}}
\frac{1}{L^{2} (L-p_{1})^{2}
\cdots (L-\sum_{i=1}^{n-1} p_{i})^{2}} \\
& = &
\frac{i (-1)^{n+1}}{\pi^{2+m-\eps}} \int
\frac{d^{4+2m}l \, d^{-2\eps}\mu}{(l^{2}-\mu^{2})
((l-p_{1})^{2}-\mu^{2})
\cdots ((l-\sum_{i=1}^{n-1} p_{i})^{2}-\mu^{2})}
\ .
\nonumber
\eeqa
The higher dimensional integral functions are related
to $4-2\eps$ dimensional integrals with a factor $\mu^{2m}$
inserted in the integrand. For $m=1,2$ one finds
\beqa
I_{n}[\mu^{2}]\equiv J_{n} & = &
(-\eps) I_{n}^{6-2 \eps}  \ ,
\\
\label{funK}
I_{n}[\mu^{4}]\equiv K_{n}  &= &
(-\eps)(1 - \eps) I_{n}^{8-2 \eps}
\, .
\eeqa

We encounter bubble functions with $m=0,1$,
triangles with one massive external line and $m=0,1$,
and boxes with four massless external lines and $m=0,1,2$:
\beqa
&& I_{2}(P^{2})\ =\
\frac{r_{\Gamma}}{\eps(1-2 \eps)} (-P^{2})^{-\eps}
\,\, , \qquad
I_{2}^{6-2 \eps}(P^{2})\ =\
-\frac{r_{\Gamma}}{2 \eps(1-2 \eps)(3-2\eps)}
(-P^{2})^{1-\eps}
\,\, , \nonumber \\
&& I_{3}(P^{2})\ =\
\frac{r_{\Gamma}}{\eps^{2}} (-P^{2})^{-1-\eps}
\qquad \,\, , \qquad
I_{3}^{6-2 \eps}(P^{2})\ =\
\frac{r_{\Gamma}}{2 \eps(1- \eps)(1-2\eps)}
(-P^{2})^{-\eps}
\,\, , \nonumber \\
&&(-\eps)I_{4}^{6-2\eps}\ =\
0 + \mathcal{O}(\eps) \,\, ,
\qquad (-\eps)(1-\eps)I_{4}^{8-2\eps}\ =\
-\frac{1}{6} + \mathcal{O}(\eps)
\,\, .
\label{ffff}
\eeqa
Note that the expressions for the bubbles and triangles are 
valid to all orders in $\e$.

\begin{figure}
\begin{center}
$
\begin{array}{lcr}
\scalebox{0.3}{\includegraphics{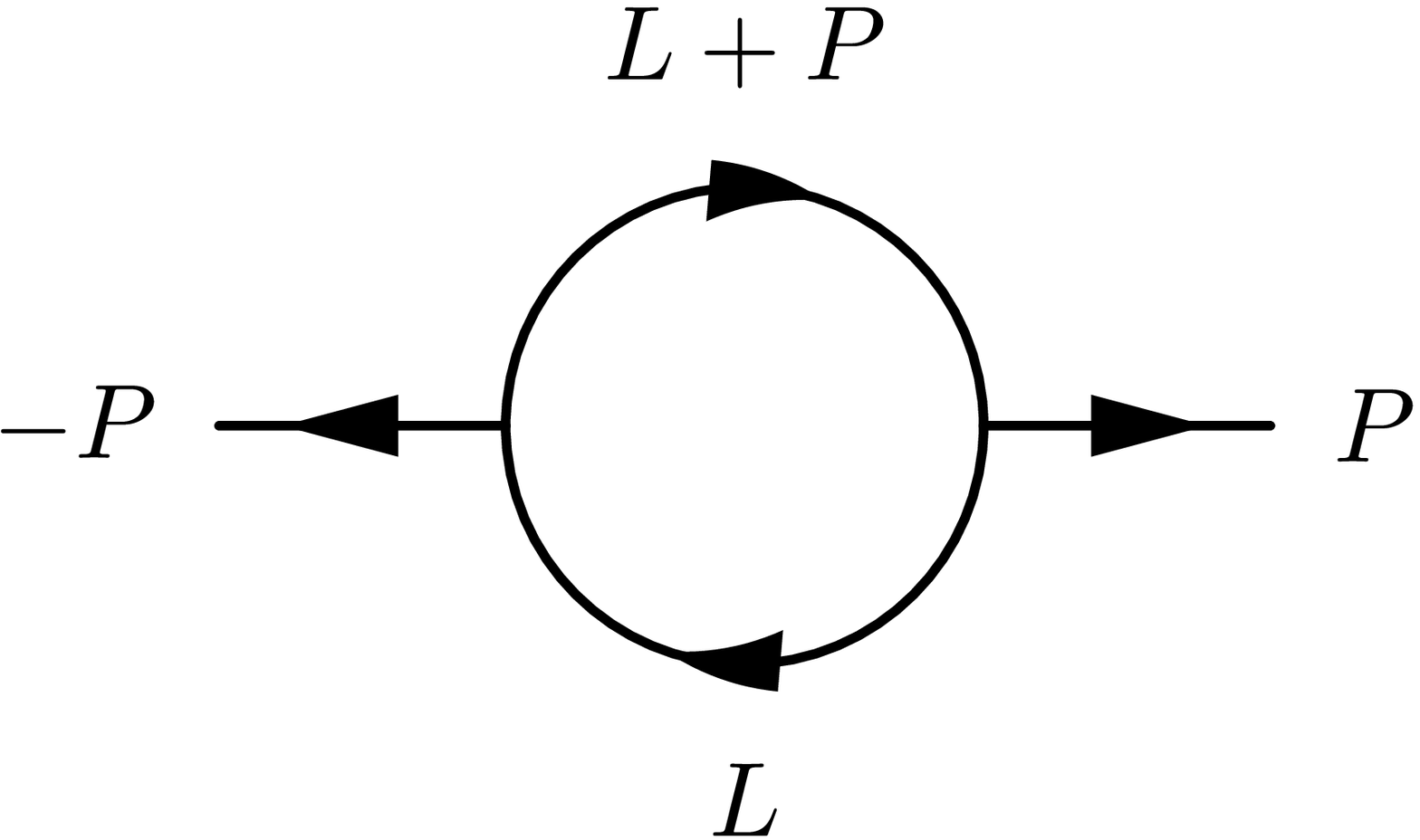}}
\label{triangle}
& \;\;\;\;\;\;\;\; &
\scalebox{0.3}{\includegraphics{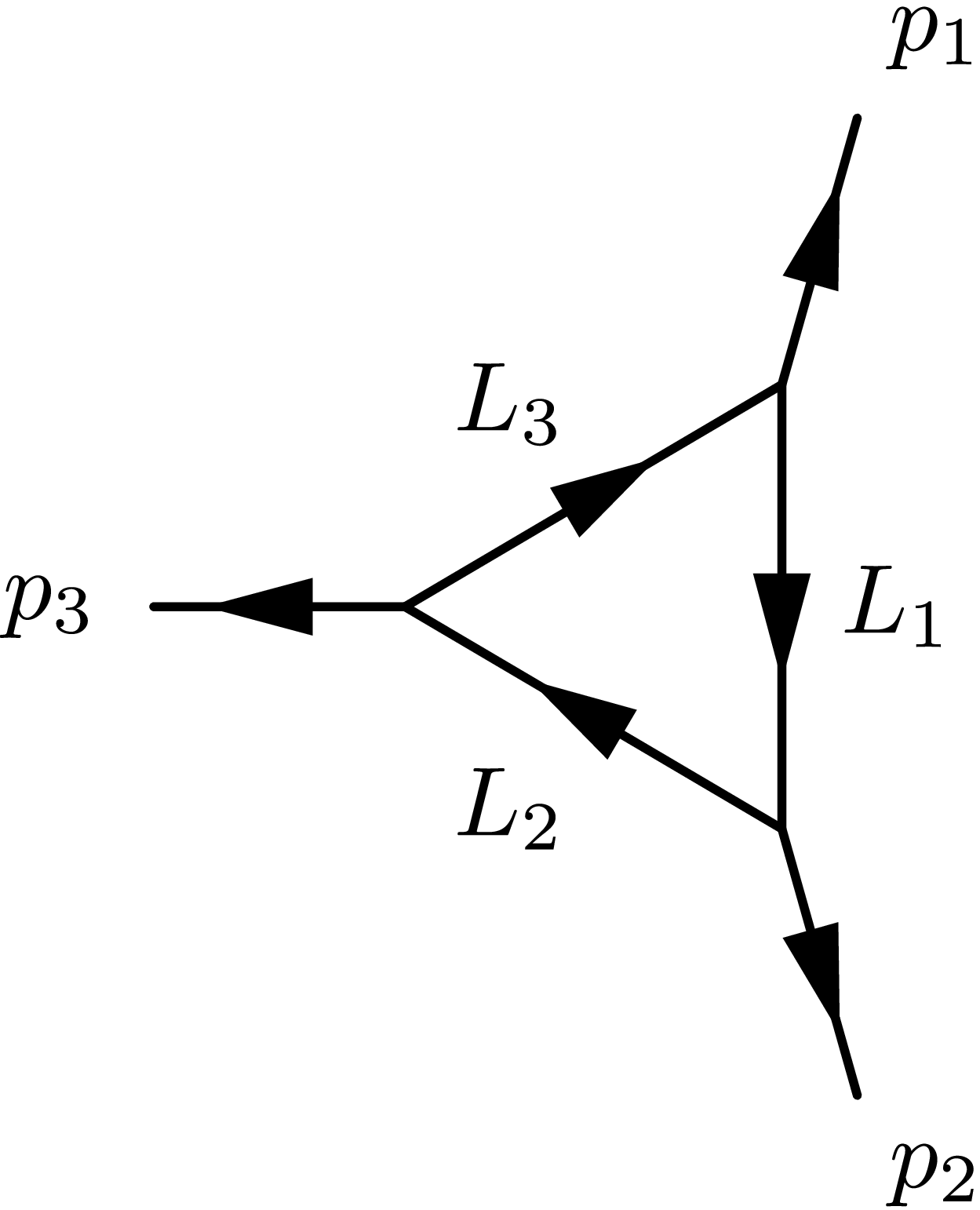}}
\end{array}
$
\caption{\it  Kinematics of the bubble and triangle integral functions
studied in this appendix.}
\end{center}
\end{figure}

We now present the result of the PV reduction for
various tensor integrals.
These are given
in terms of scalar $n$-point integral functions $I_{n}^D$
in various dimensions $D$, specifically in terms of
$I_n$, $I_n^{6-2\eps}$ and $I_n^{8-2\eps}$
in $4-2\eps$, $6-2\eps$ and  $8-2\eps$ dimensions, respectively.
The expressions are valid to all orders in $\eps$, if $I_{n}$,
$I_n^{6-2\eps}$ and $I_n^{8-2\eps}$
are evaluated to all orders, and
the PV reductions have been performed
in a fashion that naturally leads to coefficients
without explicit $\eps$ dependence
(the reader may consult  \cite{Bern:1995ix} for more details
on this particular variant of PV reductions).

For the linear and two-tensor bubbles  (Figure 4) we have 
\beqa
\label{AllPVs}
&&
I_2\big[ L^{\m} \big]
\ = \
-\frac{1}{2} I_2 P^\mu\ ,
\\ [8pt]
&&
I_2\big[ L^{\m} L^{\n}\big]
\ = \
  -\frac{1}{2} I_2^{6-2\eps}\delta^{\mu\nu}_{[4-2\eps]}
                   + \bigg(  \frac{1}{4} I_2 +
\frac{1}{2t} I_2^{6-2\eps}   \bigg) P^\mu P^\nu\ .
\eeqa

For the linear, two- and three-tensor triangles (Figure 3) we have

\beqa
&&
I_3\big[ L_{3}^{\m} \big]
\ = \
- \frac{1}{t} I_2 p_2^\mu +
                          \bigg( -I_3 +
 \frac{1}{t} I_2 \bigg) p_3^\mu\ ,
\\ [8pt]\nonumber
&&
I_3 \big[ L_{3}^{\m}L_{3}^{\n} \big]
\ = \
\frac{1}{2t} I_2 p_2^\mu p_2^\nu
                   + \bigg( \frac{1}{t} I_3^{6-2\eps} +
\frac{1}{2t} I_2 \bigg)
                   \bigg(p_2^\mu p_3^\nu+p_2^\nu p_3^\mu \bigg)
          \\ [8pt] && \qquad\qquad\qquad\qquad
           + \bigg( - \frac{3}{2t} I_2 +
I_3 \bigg) p_3^\mu p_3^\nu
                                      -\frac{1}{2} I_3^{6-2\eps} \delta^{\mu\nu}_{[4-2\eps]}\ ,
\\ [8pt]\nonumber
&&
I_3 \big[ L_{3}^{\m}L_{3}^{\n}L_{3}^{\rho} \big]
\ = \
                                   - \bigg( \frac{1}{4t} I_2 +
\frac{1}{2t^2} I_2^{6-2\eps}   \bigg)
\bigg(p_2^\mu p_2^\nu p_2^\rho \bigg)
\\ [8pt]\nonumber &&\qquad
-     \bigg(  \frac{1}{4t} I_2 +
\frac{3}{2t^2} I_2^{6-2\eps}  \bigg)
       \bigg( p_2^\mu p_2^\nu p_3^\rho +
p_2^\mu p_3^\nu p_2^\rho + p_3^\mu p_2^\nu p_2^\rho \bigg)
  \\ [8pt]\nonumber  && \qquad +
\bigg( -  \frac{1}{4t} I_2 + \frac{3}{2t^2} I_2^{6-2\eps}
- \frac{2}{t} I_3^{6-2\eps} \bigg)
\bigg(  p_2^\mu p_3^\nu p_3^\rho +
p_3^\mu p_3^\nu p_2^\rho + p_3^\mu p_2^\nu p_3^\rho \bigg)
 \\[8pt] \nonumber
&& \qquad +
\bigg(   \frac{7}{4t} I_2 + \frac{1}{2t^2} I_2^{6-2\eps}
   -  I_3 \bigg)  \bigg(  p_3^\mu p_3^\nu p_3^\rho  \bigg)
 +     \frac{1}{2t} I_2^{6-2\eps}
        \bigg(  \delta^{\mu\nu} p_2^\rho +
\delta^{\mu\rho} p_2^\nu+  \delta^{\rho\nu} p_2^\mu \bigg)
 \\ [8pt]
&& \qquad
+ \bigg(  - \frac{1}{2t} I_2^{6-2\eps}+
\frac{1}{2} I_3^{6-2\eps} \bigg)
 \bigg(  \delta^{\mu\nu} p_3^\rho +
\delta^{\mu\rho} p_3^\nu+ \delta^{\rho\nu} p_3^\mu \bigg)
\ ,
\eeqa
where momenta $p_2$ and $p_3$ are null and all integral functions appearing 
are functions of $P^2 = p_1^2=(p_2+p_3)^2$.


\newpage

\begin{thebibliography}{99}

\bibitem{witten} E.~Witten,
{\it Perturbative gauge theory as a string theory in twistor space}, Commun.\ Math.\ Phys.\
{\bf 252}, 189 (2004),  {\tt hep-th/0312171}.

\bibitem{Cachazo:2005ga}
F.~Cachazo and P.~Svr\v{c}ek, {\it Lectures on twistor strings and perturbative Yang-Mills theory},
PoS {\bf RTN2005}, 004 (2005),  {\tt hep-th/0504194}.

\bibitem{Brandhuber:2006vh}
A.~Brandhuber and G.~Travaglini,
{\it Quantum MHV diagrams}, {\tt hep-th/0609011}.

\bibitem{csw} F.~Cachazo, P.~Svr\v{c}ek and E.~Witten,
{\it MHV vertices and tree amplitudes in gauge theory}, JHEP {\bf 0409} (2004) 006, {\tt
hep-th/0403047}.

\bibitem{bst}
A.~Brandhuber, B.~Spence and G.~Travaglini, {\it One-Loop Gauge Theory Amplitudes in N=4 super
Yang-Mills from MHV Vertices}, Nucl.\ Phys.\ B {\bf 706} (2005) 150, {\tt  hep-th/0407214}.

\bibitem{quig}
C.~Quigley and M.~Rozali, {\it One-Loop MHV Amplitudes in Supersymmetric Gauge Theories}, {\tt
hep-th/0410278}.

\bibitem{bbst1}
J.~Bedford, A.~Brandhuber, B.~Spence and G.~Travaglini, {\it A Twistor Approach to One-Loop
Amplitudes in ${\cal N} \! = \! 1$ Supersymmetric Yang-Mills Theory}, Nucl.\ Phys.\ B {\bf 706}
(2005) 100, {\tt hep-th/0410280}.

\bibitem{bbst2}
J.~Bedford, A.~Brandhuber, B.~Spence and G.~Travaglini, {\it Non-supersymmetric loop amplitudes
and MHV vertices,} Nucl.\ Phys.\ B {\bf 712} (2005) 59, {\tt hep-th/0412108}.

 \bibitem{rec}
  R.~Britto, F.~Cachazo, B.~Feng and E.~Witten,
  {\it Direct proof of tree-level recursion relation in Yang-Mills theory,}
  Phys.\ Rev.\ Lett.\  {\bf 94} (2005) 181602,
  {\tt hep-th/0501052}.

\bibitem{ris}
  K.~Risager,
  {\it A direct proof of the CSW rules,}
  JHEP {\bf 0512} (2005) 003,
  {\tt hep-th/0508206}.


\bibitem{ftt}
  A.~Brandhuber, B.~Spence and G.~Travaglini,
  {\it From trees to loops and back,}
  JHEP {\bf 0601} (2006) 142,
  {\tt hep-th/0510253}.


 \bibitem{b1}
  Z.~Bern, L.~J.~Dixon and D.~A.~Kosower,
  {\it Bootstrapping multi-parton loop amplitudes in QCD,}
  Phys.\ Rev.\ D {\bf 73} (2006) 065013,
  {\tt hep-ph/0507005}.

\bibitem{b2}
  C.~F.~Berger, Z.~Bern, L.~J.~Dixon, D.~Forde and D.~A.~Kosower,
{\it Bootstrapping one-loop QCD amplitudes with general helicities,}
 Phys.\ Rev.\ D {\bf 74} (2006)  036009,
  {\tt hep-ph/0604195}.

\bibitem{b3}
  C.~F.~Berger, Z.~Bern, L.~J.~Dixon, D.~Forde and D.~A.~Kosower,
  {\it All one-loop maximally helicity violating gluonic amplitudes in QCD},
  {\tt hep-ph/0607014}.

\bibitem{z1}
  Z.~G.~Xiao, G.~Yang and C.~J.~Zhu,
{\it The rational part of QCD amplitude. I: The general formalism,}
  Nucl.\ Phys.\ B {\bf 758} (2006) 1,
  {\tt hep-ph/0607015}.


\bibitem{z2}
  X.~Su, Z.~G.~Xiao, G.~Yang and C.~J.~Zhu,
  {\it The rational part of QCD amplitude. II: The five-gluon,}
  Nucl.\ Phys.\ B {\bf 758} (2006) 35,
 {\tt hep-ph/0607016}.

\bibitem{z3}
  Z.~G.~Xiao, G.~Yang and C.~J.~Zhu,
 {\it The rational part of QCD amplitude. III: The six-gluon},
  Nucl.\ Phys.\ B {\bf 758} (2006) 53,
  {\tt hep-ph/0607017}.

\bibitem{Anastasiou:2006jv}
  C.~Anastasiou, R.~Britto, B.~Feng, Z.~Kunszt and P.~Mastrolia,
  {\it D-dimensional unitarity cut method,}
  {\tt hep-ph/0609191}.

\bibitem{pm}
  P.~Mastrolia,
  {\it On triple-cut of scattering amplitudes,}
  {\tt hep-th/0611091}.


\bibitem{Gorsky:2005sf}
  A.~Gorsky and A.~Rosly,
  {\it From Yang-Mills Lagrangian to MHV diagrams,}
  JHEP {\bf 0601} (2006) 101,
  {\tt hep-th/0510111}.

\bibitem{Mansfield}
 P.~Mansfield,
 {\it The Lagrangian origin of MHV rules,}
 JHEP {\bf 0603} (2006) 037,
{\tt hep-th/0511264}.

\bibitem{Ettle:2006bw}
  J.~H.~Ettle and T.~R.~Morris,
  {\it Structure of the MHV-rules Lagrangian,}
  JHEP {\bf 0608}, 003 (2006),
  {\tt hep-th/0605121}.

\bibitem{Bern:1995db}
  Z.~Bern and A.~G.~Morgan,
  {\it Massive Loop Amplitudes from Unitarity,}
  Nucl.\ Phys.\ B {\bf 467} (1996) 479, 
  {\tt hep-ph/9511336}.

\bibitem{bmst}
  A.~Brandhuber, S.~McNamara, B.~Spence and G.~Travaglini,
  {\it Loop amplitudes in pure Yang-Mills from generalised unitarity,}
  JHEP {\bf 0510} (2005) 011,
  {\tt hep-th/0506068}.

\bibitem{Bern:1994cg}
Z.~Bern, L.~J.~Dixon, D.~C.~Dunbar and D.~A.~Kosower, {\it Fusing gauge theory tree amplitudes
into loop amplitudes,} Nucl.\ Phys.\ B {\bf 435} (1995) 59, {\tt hep-ph/9409265}.

\bibitem{Bern:zx}
Z.~Bern, L.~J.~Dixon, D.~C.~Dunbar and D.~A.~Kosower, {\it One Loop N Point Gauge Theory
Amplitudes, Unitarity And Collinear Limits,} Nucl.\ Phys.\ B {\bf 425} (1994) 217, {\tt
hep-ph/9403226}.


\bibitem{bdkrec}
  Z.~Bern, L.~J.~Dixon and D.~A.~Kosower,
 {\it On-shell recurrence relations for one-loop QCD amplitudes,}
  Phys.\ Rev.\ D {\bf 71} (2005) 105013, 
  {\tt hep-th/0501240}.


\bibitem{Bardeen}
  W.~A.~Bardeen,
{\it Selfdual Yang-Mills theory, integrability and multiparton amplitudes,}
  Prog.\ Theor.\ Phys.\ Suppl.\  {\bf 123} (1996) 1.

\bibitem{stony}
  H.~Feng and Y.~t.~Huang,
 {\it MHV lagrangian for N = 4 super Yang-Mills},
{\tt hep-th/0611164}.

\bibitem{Thorn}
  E.~Braaten, T.~Curtright and C.~B.~Thorn,
  {\it Quantum Backlund Transformation For The Liouville Theory,}
  Phys.\ Lett.\ B {\bf 118} (1982) 115.


\bibitem{Cangemi}
  D.~Cangemi,
 {\it Self-dual Yang-Mills theory and one-loop maximally helicity violating
  multi-gluon amplitudes,}
 Nucl.\ Phys.\ B {\bf 484} (1997) 521,
{\tt hep-th/9605208}.

\bibitem{Korepin}
  V.~E.~Korepin and T.~Oota,
  {\it Scattering of plane waves in self-dual Yang-Mills theory,}
  J.\ Phys.\ A {\bf 29}, L625 (1996),
  {\tt hep-th/9608064}.


\bibitem{coleman}
S.~Coleman, {\it Soft Pions}, in {\it Aspects of Symmetry},
Cambridge University Press, 1985.


\bibitem{Mahlon1}
  G.~Mahlon,
 {\it One loop multi - photon helicity amplitudes,}
  Phys.\ Rev.\ D {\bf 49} (1994) 2197,
  {\tt hep-ph/9311213}.

\bibitem{Mahlon2}
  G.~Mahlon,
{\it Multi - Gluon Helicity Amplitudes Involving A Quark Loop,}
  Phys.\ Rev.\ D {\bf 49} (1994) 4438,
  {\tt hep-ph/9312276}.

\bibitem{csanom}
  G.~Chalmers and W.~Siegel,
  {\it Global conformal anomaly in N = 2 string,}
  Phys.\ Rev.\ D {\bf 64} (2001) 026001, 
  {\tt hep-th/0010238}.

\bibitem{Chalmers:1998jb}
  G.~Chalmers and W.~Siegel,
  {\it Simplifying algebra in Feynman graphs. II: Spinor helicity from the
  spacecone,}
  Phys.\ Rev.\ D {\bf 59} (1999) 045013,
  {\tt hep-ph/9801220}.



\bibitem{Chalmers:1996rq}
 G.~Chalmers and W.~Siegel,
 {\it The self-dual sector of {QCD} amplitudes,}
 Phys.\ Rev.\ D {\bf 54} (1996) 7628,
 {\tt hep-th/9606061}.




\bibitem{Bern:1995ix}
  Z.~Bern and G.~Chalmers,
  {\it Factorization in one loop gauge theory,}
  Nucl.\ Phys.\ B {\bf 447} (1995) 465,
  {\tt hep-ph/9503236}.


\end{thebibliography}
\end{document}